\documentclass[aps,twocolumn,prx,amsmath,amssymb,superscriptaddress]{revtex4-1}
\usepackage{multirow}
\usepackage{graphicx,epsfig}
\usepackage{color}
\usepackage{wasysym}
\usepackage{units}
\def \tr {\text{Tr}}
\providecommand{\nbraket}[1]{\left\langle#1\right\rangle}

\usepackage{dcolumn}
\usepackage{amsmath,amssymb,amsbsy}
\usepackage{amsthm}
\usepackage{graphicx,epsfig,subfigure,overpic,floatflt}
\usepackage{mathrsfs}
\usepackage{multirow}
\usepackage{hyperref}
\usepackage{url}
\usepackage{color}
\usepackage{framed}
\usepackage[sort&compress]{natbib}
\usepackage{appendix}
\usepackage{pifont}
\usepackage{comment}
\usepackage{fancyhdr}
\usepackage{listings}
\usepackage{scalefnt}
\usepackage{bbold}
\usepackage{tikz}
\usepackage[framemethod=tikz]{mdframed}

\def \beq {\begin{equation}}
\def \edq {\end{equation}}
\def \bes {\begin{subequations}}
\def \eds {\end{subequations}}
\def \beqn {\begin{equation*}}
\def \edqn {\end{equation*}}
\def \nn  {\nonumber}
\def \dag {\dagger}
\def \Up {\Uparrow}
\def \Down {\Downarrow}
\def \up {\uparrow}
\def \down {\downarrow}
\def \eps {\epsilon}
\def \sm {\sigma}
\def \bsm {\bar{\sigma}}
\def \bk {\bar{k}}
\def \bs {\bar{s}}
\def \zhup {\sqrt{\frac{2\Gamma_{\up}}{\pi}}}
\def \zhdn {\sqrt{\frac{2\Gamma_{\down}}{\pi}}}
\def \veps {\varepsilon}
\def \calu {{\cal{U}}}
\def \calh {{\cal{H}}}
\def \call {{\cal{L}}}
\def \bcalh {\bar{\cal{H}}}
\def \bcall {\widetilde{\cal{L}}}
\def \calz {{\cal{Z}}}
\def \cald {{\cal{D}}}
\def \calg {{\cal{G}}}
\def \mG {{\mathscr{G}}}
\def \bcalg {\boldsymbol{\cal{G}}}
\def \calp {{\cal{P}}}
\def \calq {{\cal{Q}}}
\def \calf {{\cal{F}}}
\def \calo {{\cal{O}}}
\def \caln {{\cal{N}}}
\def \calt {{\cal{T}}}
\def \calr {{\cal{R}}}
\def \cals {{\cal{S}}}
\def \cali {{\cal{I}}}
\def \calm {{\cal{M}}}
\def \calw {{\cal{W}}}
\def \cale {{\cal{E}}}
\def \calc {{\cal{C}}}
\def \scrh {{\mathscr{H}}}
\def \scrg {{\mathscr{G}}}
\def \scrf {{\mathscr{F}}}
\def \scrs {{\mathscr{S}}}
\def \wcalh {\widetilde{\calh}}
\def \wPsi {\widetilde{\Psi}}
\def \hatb {\widehat{b}}
\def \hatc {\hat{c}}
\def \hatd {\hat{d}}
\def \hatf {\widehat{f}}
\def \hatz {\hat{z}}
\def \hatu {\hat{u}}
\def \hatL {\hat{L}}
\def \hatV {\hat{V}}
\def \hatw {\hat{\omega}}
\def \hatg {\widehat{g}}
\def \htau {\widehat{\tau}}
\def \hatgam {\widehat{\Gamma}}
\def \hatG {\widehat{{\cal{G}}}}
\def \hatQ {\widehat{{\cal{Q}}}}
\def \hatcP {\widehat{{\cal{P}}}}
\def \hatsig {\widehat{\Sigma}}
\def \hatsm {\widehat{\sigma}}
\def \bsig {{\mathbf{\Sigma}}}
\def \bV {{\mathbf{V}}}
\def \bfsm {{\boldsymbol{\sm}}}
\def \wtsig {{\widetilde{\Sigma}}}
\def \talpha {\widetilde{\alpha}}
\def \tell {\widetilde{\ell}}
\def \tmu {\widetilde{\mu}}
\def \tDelta {\widetilde{\Delta}}
\def \tp {\widetilde{p}}
\def \ty {\widetilde{y}}
\def \tz {\widetilde{z}}
\def \tPi {\widetilde{\Pi}}
\def \teps {\widetilde{\eps}}
\def \tom {\widetilde{\omega}}
\def \tgamma {\widetilde{\gamma}}
\def \tr {\text{Tr}}
\def \det {\text{det}}
\def \bo {\bf{1}}
\def \bt {\bar{t}}
\def \bp {\bar{p}}
\def \bq {\bar{q}}
\def \balpha {\bar{\alpha}}
\def \lAngle {\langle\langle}
\def \rAngle {\rangle\rangle}
\def \bzero {\tilde{0}}
\def \coffee {{\Huge{\Coffeecup}}}
\def \stop {{\Huge{\Stopsign}}}
\def \keyboard {{\Huge{\Keyboard}}}
\def \tGamma {\widetilde{\Gamma}}
\def \tPsi {\tilde{\Psi}}
\def \tg {\hat{\text{g}}}
\def \ttg {\text{g}}
\def \bttg {\text{$\mathbf{g}$}}
\def \mttg {\text{\bf{g}}}
\def \wtV {\widetilde{V}}
\def \wtt {\widetilde{t}}
\def \wtz {\widetilde{z}}
\def \tb {\widetilde{b}}
\def \tc {\tilde{c}}
\def \hatt {\hat{t}}
\def \bA {\mathbf{A}}
\def \bX {\mathbf{X}}
\def \bY {\mathbf{Y}}
\def \bnu {\boldsymbol{\nu}}
\def \matone {\mathbb{1}}
\def \Pf {\text{Pf}}
\def \whand {\huge{\Writinghand}}
\def \hc {\text{H.c.}}

\providecommand{\abs}[1]{\lvert#1\rvert}
\providecommand{\norm}[1]{\lVert#1\rVert}
\providecommand{\brakets}[2]{\langle#1|#2|#1\rangle}
\providecommand{\Brakets}[3]{\langle#1|#2|#3\rangle}
\providecommand{\nbraket}[1]{\left\langle#1\right\rangle}
\providecommand{\dbraket}[1]{\langle\langle#1\rangle\rangle}

\begin{document}
\title{Shiba states and zero-bias anomalies in the hybrid normal-superconductor Anderson model}
\author{Rok \v{Z}itko}
\affiliation{Jo\v{z}ef Stefan Institute, Jamova 39, SI-1000 Ljubljana,
Slovenia}
\affiliation{Faculty  of Mathematics and Physics, University of Ljubljana,
Jadranska 19, SI-1000 Ljubljana, Slovenia}
\author{Jong Soo Lim}
\affiliation{School of Physics, Korea Institute for Advanced Study, Seoul 130-722, Korea}
\author{Rosa L{\'o}pez}
\affiliation{Instituto de F\'{\i}sica Interdisciplinar y Sistemas Complejos
IFISC (UIB-CSIC), E-07122 Palma de Mallorca, Spain}
\affiliation{Kavli Institute for Theoretical Physics, University of California, Santa Barbara, California 93106-4030, USA}
\author{Ram\'on Aguado}
\affiliation{Instituto de Ciencia de Materiales de Madrid, Consejo
Superior de Investigaciones Cient\'{\i}ficas (ICMM-CSIC), Sor Juana
In\'es de la Cruz 3, 28049 Madrid, Spain}

\begin{abstract} Hybrid semiconductor-superconductor systems are interesting melting pots where various fundamental effects in condensed matter physics coexist. For example, when a quantum dot is coupled to a superconducting electrode two very distinct phenomena, superconductivity and the Kondo effect, compete. As a result of this competition, the system undergoes a quantum phase transition when the superconducting gap $\Delta$ is of the order of the Kondo temperature $T_K$. The underlying physics behind such transition ultimately relies on the physics of the Anderson model where the standard metallic host is replaced by a superconducting one, namely the physics of a (quantum) magnetic impurity in a superconductor. A characteristic feature of this hybrid system is the emergence of sub-gap bound states, the so-called Yu-Shiba-Rusinov (YSR) states, which cross zero energy across the quantum phase transition, signaling a switching of the fermion parity and spin (doublet or singlet) of the ground state. Interestingly, similar hybrid devices based on semiconducting nanowires with spin-orbit coupling may host exotic zero-energy bound states with Majorana character. Both, parity crossings and Majorana bound states (MBS)s, are experimentally marked by zero bias anomalies in transport, which are detected by coupling the hybrid device with an extra normal contact. We here demonstrate theoretically that this extra contact, usually considered as a non-perturbing tunneling weak probe, leads to nontrivial effects. This conclusion is supported by numerical renormalization group calculations of the phase diagram of an Anderson impurity coupled to both superconducting and normal-state leads. We obtain this phase diagram for an arbitrary ratio $\Delta\over T_K$ for the first time, which allows us to analyze relevant experimental scenarios, such as parity crossings as well as novel Kondo features induced by the normal lead, as this ratio changes. Spectral functions at finite temperatures and magnetic fields, which can be directly linked to experimental tunneling transport characteristics, show zero-energy anomalies irrespective of whether the system is in the doublet or singlet regime. We also derive the analytical condition for the occurrence of Zeeman-induced fermion-parity switches in the presence of interactions which bears unexpected similarities with the condition for emergent MBSs in nanowires.
\end{abstract}

\maketitle

\section{Motivation and Introduction}
\begin{figure*}
\centering
\includegraphics[width=1\textwidth]{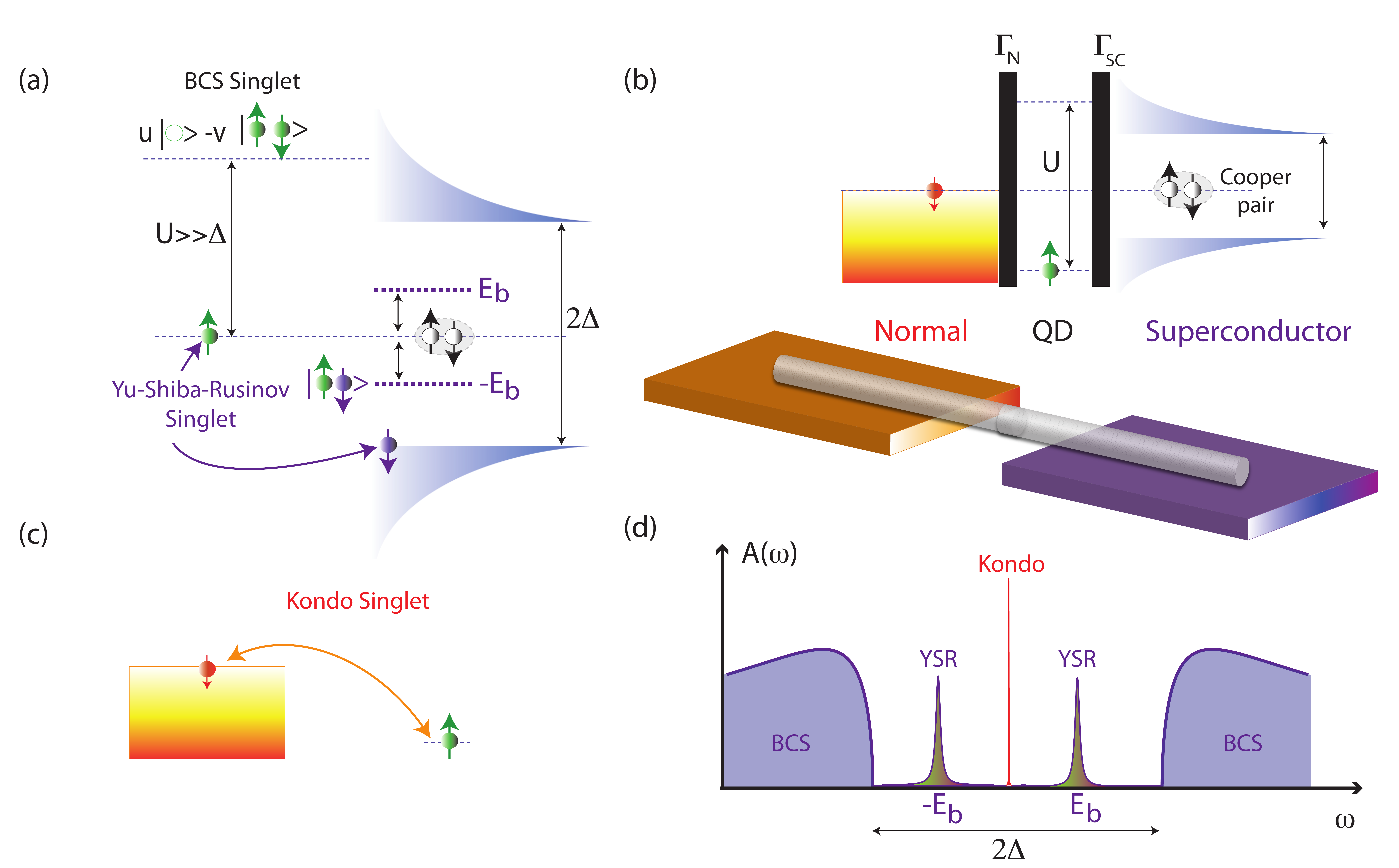}
\caption{(Color online) (a) Lowest energy many-particle eigenstates of
an Anderson impurity coupled to a superconductor with the typical BCS
density of states $\sim [(\omega/\Delta)^2-1]^{-1/2}$ for large
on-site interaction $U \gg \Delta$. The magnetic impurity ground state
develops singlet correlations with the quasiparticles in the
superconducting leads and forms a Yu-Shiba-Rusinov-like (YSR) singlet
eigenstate.
This excited state gives rise to subgap spectral peaks at energies $E_b$
and $-E_b$. When these subgap excitations cross zero energy, the
system undergoes a parity-changing quantum phase transition and the YSR singlet
becomes the new ground state. At higher energies there are BCS-like
excited singlet states resulting from the hybridization between the empty
and doubly occupied states of the quantum impurity. These singlets 
occur at subgap energies in the opposite limit $U \ll \Delta$ (not
shown). (b) Top: Schematics of a normal-quantum dot-superconducting
hybrid system with all the relevant energies involved in the problem.
In odd-occupancy Coulomb blockade valleys (charging energy $U$), the unpaired
spin (green) mimics the physics of a magnetic impurity coupled to a
superconductor (coupling $\Gamma_{SC}$) with a BCS density of states
(purple) with gap $\Delta$. This physics can be considerably modified
by the weak coupling ($\Gamma_{N}$) to a normal probe (orange-yellow),
as we discuss in this work. Bottom: this hybrid system can be realized with,
e. g., nanowires deposited on top of normal and superconducting
electrodes.  (c) Standard Kondo singlets that occur as quasiparticles
in the normal metal (red) screen the magnetic doublet. (d) Typical
spectral density of the hybridized quantum dot in the magnetic doublet
ground state regime showing the coexistence of YRS singlet subgap
excitations and a Kondo resonance. The subgap excitations remove
spectral weight from the BCS density of states.}
\label{fig1a}
\end{figure*}
The Kondo effect has been fundamental in furthering our understanding of strong correlations in condensed matter physics. First observed some 80 years ago \textcolor{blue}{\cite{Haas34}},  the anomalous behavior of the low-temperature resistivity of dilute magnetic alloys can be understood as the many-body screening of magnetic moments in a metal. This screening occurs via quasiparticle spin exchange well below the Kondo temperature $T_K$ \textcolor{blue}{\cite{Kondo64a,Hewson}}. 
During the last decades the interest in the Kondo effect has revived following its discovery in quantum dots based on
semiconductors \textcolor{blue}{\cite{KondoreviewQD}},  carbon nanotubes \textcolor{blue}{\cite{KondoCNT}} and nanowires \textcolor{blue}{\cite{KondoNWs}}. Quantum dots behave as magnetic impurities but, in contrast to real ones, are fully tunable such that Kondo physics can be controlled in precise detail. 

Interestingly, hybrid devices based on quantum dots coupled to
superconductors can also be fabricated and the physics of magnetic
impurities in a superconductor can be studied in an unprecedented
manner \textcolor{blue}{\cite{Silvanoreview}}. A characteristic feature of these systems
is the presence of sub-gap excitations, the so-called Yu-Shiba-Rusinov
(YSR) bound states or simply Shiba states \textcolor{blue}{\cite{subgap,Balatsky06}},
that appear owing to the pair-breaking effects that magnetic moments
have on superconductivity. Their physical meaning can be understood
already at the level of a classical spin $S$ exchange-coupled to the
superconductor by a coupling $J$. This interaction gives rise to an
effective magnetic field $JS$ which lowers the energy for
quasiparticle excitations by an amount:
\begin{equation}
E_b=\Delta\frac{1-(\pi JS\rho_0)^2}{1+(\pi JS\rho_0)^2},
\label{eq:Eb}
\end{equation}
where $\rho_0$ is the normal state density of states at the Fermi
energy and $\Delta$ is the superconducting gap. For weak exchange,
$JS\ll1/\pi\rho_0$, the ground state is a standard BCS wave function,
with all single particle states forming Cooper pairs, plus an
unscreened impurity spin. Single quasiparticle excitations on top of
this ground state, as described by Eq.~\textcolor{blue}{\eqref{eq:Eb}}, occur at energies close to
the gap. For large enough $J$, however, $E_b$ can cross zero energy
such that the state with one unpaired quasiparticle, which is a
non-BCS state, becomes the new ground state. Zero energy crossings of
the YRS state thus signal a quantum phase transition  (QPT)  where  \emph{the fermionic parity of the ground state changes} \textcolor{blue}{\cite{Sakurai}}. 

Quantum fluctuations lead to a very complex scenario since exchange is mediated by Kondo processes. In a superconductor
no quasiparticles are available below the gap $\Delta$, hence Kondo screening is incomplete. To analyze all possible ground states, let us consider a single, spin-degenerate quantum impurity level coupled to a superconductor.  In general, two spin states are possible: a spin doublet (spin 1/2), $|D\rangle=\uparrow,\downarrow$ and a spin singlet (spin zero), $|S\rangle$. The latter can be of two types (apart from the standard Cooper pairs of the BCS ground state): Kondo-like superpositions between the spin doublet and Bogoliubov quasiparticles in the superconductor and BCS-like superpositions of zero and doubly occupied states of the impurity level (Fig.~\textcolor{blue}{\ref{fig1a}(a)}).  In the weak Kondo coupling regime ($T_K\ll\Delta$), the ground state is the doublet while Kondo-like singlet excitations give rise to YSR bound states (assuming large on-site interaction $U\gg\Delta$, such that the BCS-like singlets are higher in energy than the Kondo ones, Fig.~\textcolor{blue}{\ref{fig1a}(a)}). The position in energy of these YSR excitations smoothly evolves from $E_b \simeq \Delta$ towards 
positions close to the Fermi level when $T_K\sim\Delta$. At larger $T_K$, the YRS cross zero energy and the system undergoes a parity-changing QPT where the new ground state is now the Kondo singlet \textcolor{blue}{\cite{Sakai93}}.

Experimentally, these complicated correlations can be determined by the transport
spectroscopy of a quantum dot (QD) coupled to, both, a superconductor and a weak normal lead (Fig.~\textcolor{blue}{\ref{fig1a}(b)}). Sub-gap features in the differential conductance of this setup
can be directly ascribed to YSRs
\textcolor{blue}{\cite{Jespersen,Eichler,Grove-RasmussenPRB:09,Pillet10,Deacon10a,Dirks11,AndersenPRL:11,Lee12,Kim13,Chang13,Rok13,Kumar14,Schindele14,Lee14}}.
Zero bias anomalies (ZBA)s, in particular, mark QPT parity crossings
\textcolor{blue}{\cite{Deacon10a,Franke11,Lee14}}.


More recently, subgap states have attracted a great deal of attention in the context of topological superconductors containing Majorana bound states (MBS). These MBS are far more elusive than standard YRS and were predicted to appear as zero-energy bound states in effective 
spinless p-wave nanostructures, such as the ones resulting 
from the combined action of spin-orbit coupling and Zeeman 
splitting in nanowires proximized with s-wave superconductors
\textcolor{blue}{\cite{Reviews}}. These nanowire devices, very similar to the ones where
the YSR parity crossings have been reported, see, e. g., Refs. [\textcolor{blue}{\onlinecite{Chang13}}] or [\textcolor{blue}{\onlinecite{Lee14}}], are expected to become topological superconductors
when the following criterion is satisfied \textcolor{blue}{\cite{Lutchyn:PRL10,Oreg:PRL10}}
\begin{equation}\label{criterion}
E_Z^2=\Delta_{*}^2+\mu^2
\end{equation}
where $E_Z=g\mu_B B/2$ is the Zeeman energy ($g$ is the $g$-
factor and $\mu_B$ is the Bohr magneton), $\Delta_{*}$ is the
proximity-induced superconducting pairing, and $\mu$ is the 
chemical potential. Indeed, recent experiments have reported
ZBAs in transport through proximized nanowires that can be interpreted as signatures of Majorana
states \textcolor{blue}{\cite{Mourik12,Das12ZBA,Deng12,Finck:PRL13,Churchill:PRB13}}. Alternative explanations
involving Kondo physics and the associated YSR states were dismissed
based on the expected shifts with increasing magnetic field $B$. 

As we will discuss in
this work, however, the interplay of strong Coulomb interaction,
Zeeman splitting, as well as the hybridization to the normal-state
tunneling probe, leads to unanticipated manifestations of Kondo
physics, similar to the signatures of Majorana states. For YSR and MBS
alike, the zero-bias anomalies can be induced by the magnetic field and split into two peaks under certain circumstances. For MBS, the
field plays the crucial role of rendering the system effectively
spinless \textcolor{blue}{\cite{Lutchyn:PRL10,Oreg:PRL10}}, while the subsequent
splitting could be due to finite-size effects \textcolor{blue}{\cite{long}}. For YSRs, the field can induce
parity crossings in two ways: through the Kondo effect (by reducing the gap
so that $\Delta(B)\lesssim T_K$ \textcolor{blue}{\cite{Lee12}}) or via Zeeman splitting
of YSRs \textcolor{blue}{\cite{Lee14}}. The analysis is additionally complicated by the
presence of the tunneling probe which not only trivially broadens the
sub-gap bound states into resonances of finite width, but also leads
to further Kondo screening.


Interestingly, it has been shown \textcolor{blue}{\cite{Stanescu:PRB13}} that
Zeeman-induced crossings in very short quantum-dot-like noninteracting
nanowires smoothly evolve towards the true MBS as the wire becomes
longer. Along similar lines, recent proposals have discussed the
possibility of obtaining MBSs in chains of magnetic atoms deposited on
top of superconducting surfaces \textcolor{blue}{\cite{chain}}. In such proposals, the
YSR bound states on each impurity overlap considerably and form a
Shiba band along the chain. Remarkably, this Shiba band can support a
topological phase with end MBS, which is yet another example where YSR
bound states smoothly evolve towards MBS. The recent experimental
observations reported in Ref. \textcolor{blue}{\cite{Nadj14}} using spatially resolved
scanning tunneling spectroscopy reveal the existence of nearly
zero-energy quasi-bound energy states that, however, are too localized
to be reconciled with the Shiba band picture of Majorana end states. A
recent theoretical work \textcolor{blue}{\cite{arXiv:1412.0151}} considers a linear
chain of Anderson impurities on a superconductor as the minimal model
that might explain the strong localization. While the above works
suggest an interesting connection between the physics of magnetic
impurities in superconductors and MBS, they neglect quantum
fluctuations (and hence Kondo physics), which are essential for a
proper understanding of the YSR bound states.

This state of affairs motivates a detailed study of the minimal Anderson model incorporating both
superconducting lead and normal-state tunneling probe, and fully taking into account quantum fluctuations for an arbitrary ratio of
the gap to the Kondo temperature.  While many theoretical papers have already studied transport 
in normal-quantum dot-superconductor system \textcolor{blue}{\cite{Rodero11,KoertingPRB:10,Yamada:PRB11,BaranskiJPC:13,Zapalska:13}}, 
the precise role that the coupling $\Gamma_N$ to the normal lead has on the phase diagram
(beyond trivial broadening effects) remains largely unknown. The
presence of the tunneling probe not only trivially broadens the
sub-gap bound states into resonances of finite width, but also leads
to further Kondo screening that generates an additional spectral peak
pinned to zero frequency. 

\begin{figure*}
  \centering
\includegraphics[width=0.8\textwidth]{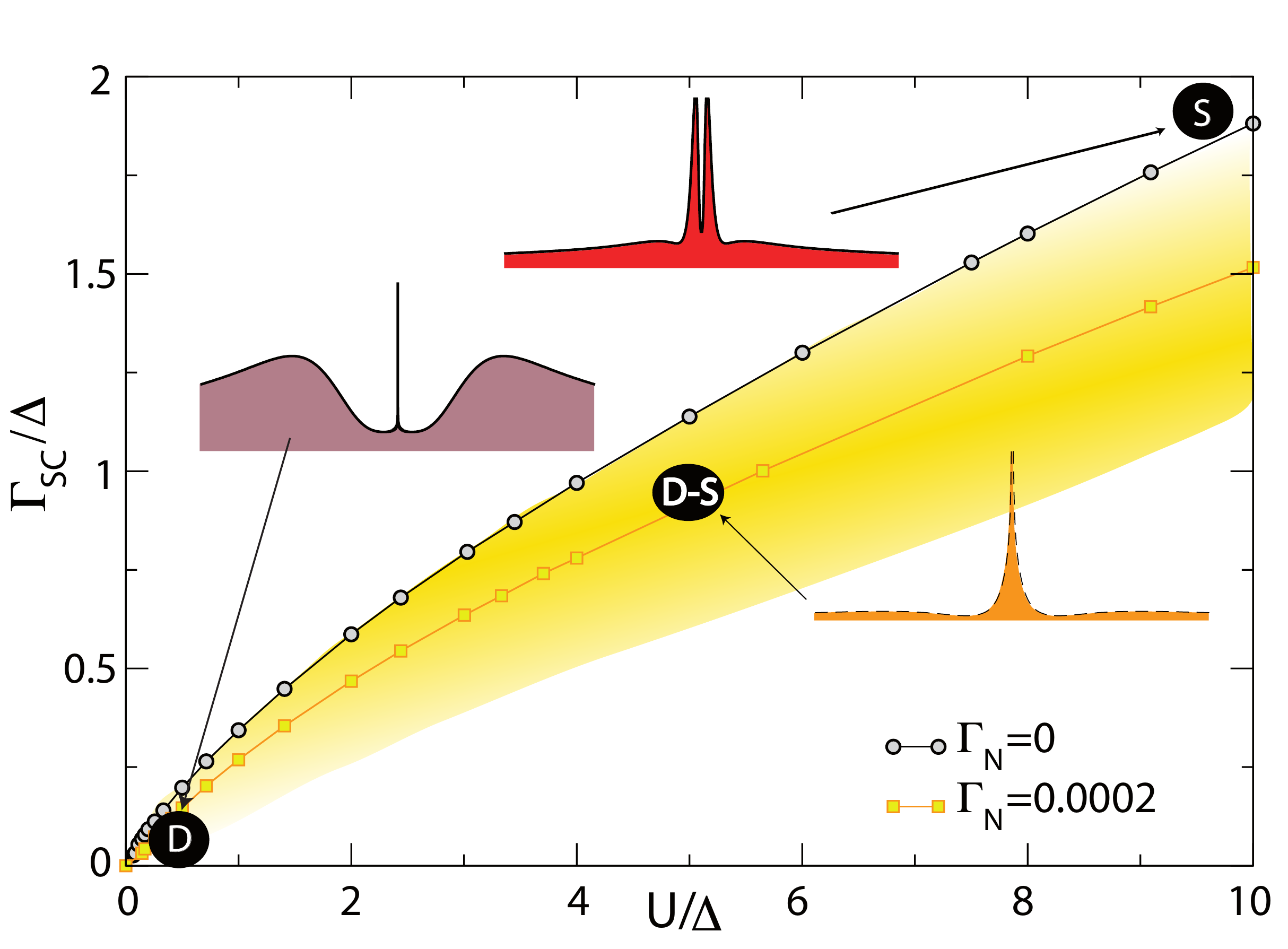}
\caption{(Color online) Phase diagram for fixed $U=0.01$ and
typical spectra for doublet (D), singlet (S) and doublet-singlet
cross-over (DS) regimes. Shading indicates the estimated width of the
cross-over region.}
\label{fig2}
\end{figure*}

To address the investigation of the YSR subgap states in this
minimal hybrid normal-superconductor Anderson model, we employ a sophisticated and almost \textit{exact} theoretical technique: 
\textit{the numerical renormalization group } (NRG) \textcolor{blue}{\cite{Wilson2,Krishna,Hofstetter,Bulla}}. The only NRG calculations of the system studied here were performed
in the $\Delta\rightarrow\infty$ limit
\textcolor{blue}{\cite{Tanaka:JPSJ07,Oguri:PRB13}}, which is unsuitable
for understanding realistic experimental situations (arbitrary ratios
$\Delta/T_K$) since they exclude all effects of the quasiparticles in the
superconductor.
We discuss the equilibrium properties  of hybrid QD systems such as
the local density of states of the quantum dot that provides useful
information for the interpretation of experimental findings for the nonlinear conductance \textcolor{blue}{\cite{Das12ZBA,Deng12,Finck:PRL13,Churchill:PRB13}}.  Some of our main
results are summarized in Fig.~\textcolor{blue}{\ref{fig2}}. Weak coupling to the normal lead,
usually considered to be just a non-perturbing tunneling probe that
may be ignored in the calculations, changes the phase diagram
considerably by replacing the sharp doublet-singlet quantum phase
transition line with a very broad cross-over region with properties
intermediate between those in the respective limits. The spectral
functions exhibit a rich phenomenology with zero bias anomalies of
different origins. In the doublet (D) regime, where the impurity would
remain unscreened for $\Gamma_N=0$ down to zero temperature, there is
a needle-like resonance due to a Kondo effect with very low Kondo
temperature $T_K^N \ll T_K$, which may already have been observed
\textcolor{blue}{\cite{Chang13}}.  Here $T_K^N$ is the Kondo temperature associated with
the screening from the weakly coupled normal-state lead, while $T_K$ is
the standard Kondo temperature associated with the screening from the
strongly-coupled superconducting lead. During the doublet-singlet (DS)
cross-over the Shiba resonances merge with the needle Kondo peak to
produce an enhanced ZBA of large amplitude. In the singlet (S) regime,
this resonance splits into two Shiba states and there is no
needle-like feature. In this regime, the magnetic field induces
further ZBA through Zeeman splitting of the doublet YSR state, see
Fig.~\textcolor{blue}{\ref{fig8}}. We derive the analytical condition for the occurrence of these Zeeman-induced fermion-parity switches in the presence of interactions. Interestingly,  the equation describing these fermion-parity switches, Eq.~\textcolor{blue}{\eqref{eq2}}, bears unexpected similarities to the inequality for MBS formation in nanowires \textcolor{blue}{\eqref{criterion}}.

This work is structured as follows. In Sec.~II we describe the model
and provide some details about the numerical technique. In Sec.~III we
present the results for the modifications of the phase diagram induced
by the normal-state lead. In Sec.~IV we discuss the effect of finite
temperatures and in Sec.~V those of the external magnetic field. Apart from NRG numerical results, this section also contains an analytical derivation of the condition for Zeeman-induced parity crossings in the presence of interactions. 
Some additional technical details are provided in the Appendices. They include a detailed discussion about the definition of the cross-over lines in the phase diagram (Appendix A) and a Schrieffer-Wolff transformation including both normal and superconducting leads (Appendix B).

\section{Model and method}

The physical system under consideration is a nanodevice (such as a
segment of a nanowire) where charge can be trapped under the effect of
electric potentials. If the number of confined electrons is small,
such that the separation between the energy levels is non-negligible,
the device can be considered as a quantum dot. In the simplest case,
there will be a single orbital. This orbital hybridizes with a
superconducting substrate as well as with a tunneling probe, and it is
exposed to an external magnetic field. We thus consider the following Anderson
impurity model (see the schematic representation in
Fig.~\textcolor{blue}{\ref{fig1a}(b)})
\begin{widetext}
\begin{equation}
H = 
\xi (n_d-1) +\frac{U}{2} (n_{d}-1)^2 + g\mu_B B S_z 
+ \sum_{k,\sigma,\alpha} \epsilon_{k_\alpha}c_{k_\alpha\sigma}^\dagger c_{k_\alpha\sigma} 
+ \sum_{k,\sigma,\alpha} \left(V_{\alpha }d^\dagger_\sigma c_{k_\alpha\sigma}+h.c.
\right)
+ \sum_k \left(\Delta c_{k_{SC}\uparrow}^\dagger
c_{k_{SC}\downarrow}^\dagger +h.c.\right).
\end{equation}
\end{widetext}
$c^\dagger$ creates an electron in the normal or superconducting lead
($\alpha=\{ N,SC \}$ is the channel index) and $d^\dagger$ at the
impurity level. The impurity occupation is
$n_d=n_{d\uparrow}+n_{d\downarrow}$ with $n_\sigma=d_\sigma^\dagger
d_\sigma$, while its spin is $S_z=(n_\uparrow-n_\downarrow)/2$. The
parameter $\xi\equiv \epsilon+\frac{U}{2}$, where $\epsilon$ is the
impurity level and $U$ the on-site repulsion, measures deviations from
the particle-hole symmetry when the occupancy is fixed exactly at 1.
Here, for simplicity, we shall focus on electron-hole symmetric
configurations $\xi=0$, unless stated otherwise. The coupling between
the impurity and the leads is described by the amplitudes $V_{\alpha}$
which define two tunneling rates: $\Gamma_{\alpha}=\pi
|V_{\alpha}|^2\rho_{\alpha}$, where $\rho_\alpha$ is the density of
states of the lead.  The energy unit is half the bandwidth. The
Hamiltonian does not include any spin-orbit coupling which is known
not to qualitatively affect Kondo physics because it does not break
the Kramers degeneracy
\textcolor{blue}{\cite{meir1994,zitko2010,malecki2007,zitko2011}}.

Since we are aiming at an accurate non-perturbative study of the
problem, we adopt the NRG method
\textcolor{blue}{\cite{Wilson2,Krishna,Hofstetter,Bulla}}. The NRG is essentially an
exact diagonalization procedure where the only approximations are the
discretization of the continuum of states in the leads, and the
truncation of the almost decoupled high-energy excitations at each
iteration step; both are controlled and, in principle, accuracies
below 1\% can be achieved. The calculations become numerically
demanding as the number of ``channels'' (i.e., leads, here one normal
and one superconducting) increases and as the symmetry is reduced
(here the only remaining symmetry in the presence of the magnetic
field is the conservation of the spin projection $S_z$). The present
problem is at the very boarder of the currently feasible NRG
computations. We employ an iterative diagonalization scheme which
consists of including a single site from the Wilson chains in each NRG
step, alternatively from the superconducting and from the normal-state
lead; we have verified that the difference compared to the
conventional approach where two sites are included at once is
inconsequential (differences of a few percent). Here this approach
works very well because the two channels have very asymmetric coupling
and are different in nature, thus the alternating site adding does not
lead to the breaking of the energy-scale separation that is necessary
in the NRG approach. The discretization parameter was $\Lambda=4$ and
we typically kept up to 6000 multiplets per NRG iteration. We made use
of the spin symmetry: SU(2) in the absence of field, U(1) in its
presence. The spectral function is calculated using the full-density
matrix algorithm which is the most reliable approach at finite
temperatures \textcolor{blue}{\cite{weichselbaum2007}}.

All relevant physical quantities can be extracted from the QD Green's
functions in the Nambu space defined as 
\begin{equation}
\hat{G}(t,t') = -i \langle
\Psi(t)\Psi^{\dagger}(t')\rangle,
\end{equation}
where $\Psi= ( d_{\uparrow} \; d^{\dagger}_{\downarrow})^T$. The
spectral function $A(\omega)$ is defined as 
\begin{equation}
A(\omega)=-\frac{1}{\pi} \mathrm{Im}
G^r_{dd}(\omega),
\end{equation}
where $G^r_{dd}(\omega)$ is the Fourier transform
of the QD retarded Green's function, namely
\begin{equation}
G^r_{dd}(\omega)=-i\int_0^\infty \mathrm{d} t\, e^{i\omega t}\langle
\{d_\sigma(t),d^\dagger_\sigma(0)\}\rangle.
\end{equation}
The doublet-singlet transition can be characterized by the changes in the
anomalous spectral function 
\begin{equation}
B(\omega)=-\frac{1}{\pi} \mathrm{Im}
F^r_{dd}(\omega)
\end{equation}
of the anomalous component of the propagator
\begin{equation}
F^r_{dd}(\omega)=-i\int_{0}^\infty \mathrm{d}t \, e^{i\omega
t}\langle \{d_\uparrow(t),d_\downarrow(0)\} \rangle.
\end{equation}

For computing spectral functions we performed averaging over $N_z=8$
interleaved discretization grids. Since the impurity is coupled to
both normal-state and superconducting channels, we performed the
broadening using a standard log-Gaussian scheme with $b=0.6$.

\section{Phase diagram}
For $\Gamma_N=0$, large $U$ favors a doublet ground state: in the
analytically solvable $\Delta\rightarrow\infty$ limit, the doublet
phase occurs for $\Gamma_{SC}$ below the line
\begin{equation}
U=2\sqrt{\xi^2+\Gamma_{SC}^2}.
\end{equation}
For finite $\Delta$, the DS transition needs to be computed
numerically (black line with circles in Fig.~\textcolor{blue}{\ref{fig2}}). Large $\Delta$ favors
a superconducting singlet state, while for smaller $\Delta$ Kondo
correlations mediated by quasiparticles above the superconducting gap
are also possible and the singlet becomes predominantly of Kondo
character as $\Gamma_{SC}$ increases. In this section we discuss how
this picture is modified by the presence of the normal-state lead. We
describe different criteria for identifying the doublet-singlet
cross-over region, the origin of the additional zero-bias anomalies, and
provide numerical results for the $\Gamma_N$ dependence.

\subsection{Phase transition vs. cross-over behavior}

For $\Gamma_N\neq0$, Kondo screening leads to a singlet ground state
for all parameter values. We emphasise that this is a statement about
the true zero-temperature ground state and that the characteristic
temperature scale for reaching such a ground state can be
exponentially low, thus experimentally irrelevant. In such
circumstances, it is more important to understand the properties at
intermediate experimentally relevant temperature scales. We find that
the sharp DS quantum phase transition for $\Gamma_N \to 0$ is replaced
at $\Gamma_N \neq 0$ by a smooth cross-over between the ``singlet''
and ``doublet regimes'' which can be empirically distinguished by
analogy with the $\Gamma_N=0$ case in several ways:
\begin{enumerate}
\item[(a)] sign of the local pairing term $\langle d_\uparrow d_\downarrow
\rangle$;

\item[(b)] merging and splitting of Shiba resonances in the regular
spectral function $A(\omega)$;

\item[(c)] peak weights in the anomalous spectral function $B(\omega)$.
\end{enumerate}
These criteria are fully equivalent for $\Gamma_N=0$ when the DS
transition marks a true discontinuity in all physical properties, but
they define three different lines for finite $\Gamma_N$ because the
cross-over is smooth and extended. The line with squares in
Fig.~\textcolor{blue}{\ref{fig2}} corresponds to criterion a. The width of the
cross-over region, indicated by the shading in Fig.~\textcolor{blue}{\ref{fig2}},
roughly indicates the range where the YSR resonances are merged
(criterion b, which is experimentally the most relevant). Due to the
significant width of the cross-over region even for small $\Gamma_N$,
the normal-state electrode cannot be considered as a non-perturbing
probe.

Further details about the conceptual and technical issues related to
defining the position of the cross-over lines are given in Appendix~A.

\subsection{Origin of the zero bias anomalies}

\begin{figure}
  \centering
\includegraphics[width=0.5\textwidth]{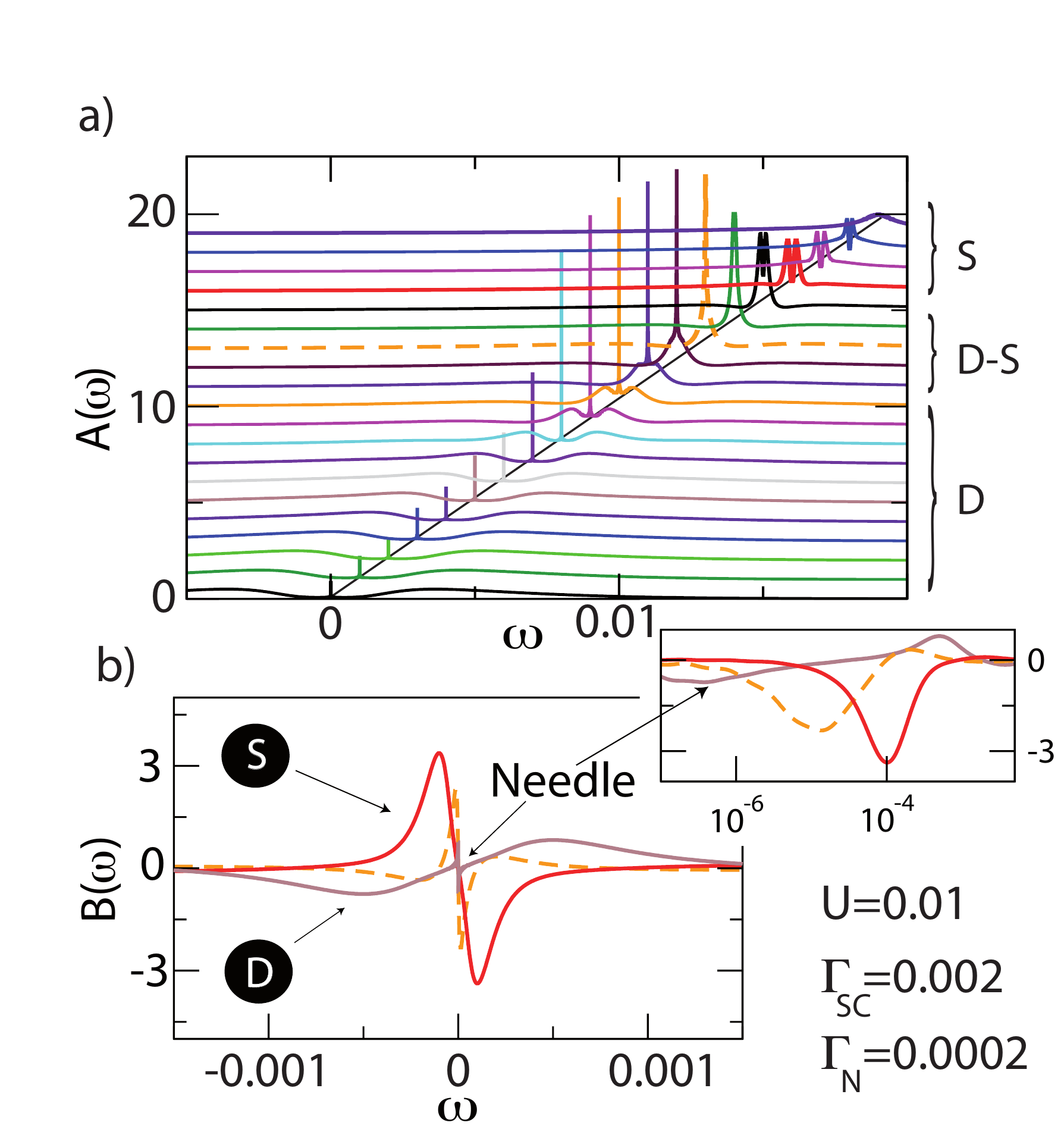}
\caption{(Color online) a) Spectra for $\Delta$ ranging from $0.5$ (bottom) to $0$ (top).
Offsets are added for clarity.
%
b) Anomalous spectral function $B(\omega)$ for $\Delta=0.004$
(doublet), $\Delta=0.002$ (doublet-singlet cross-over, dashed line)
and $\Delta=0.001$ (singlet). Inset: $B(\omega)$ for $\omega>0$ on the
logarithmic frequency scale. The arrow indicates the peak with
negative weight in the doublet regime, which is associated with the
Kondo effect and the ultimate spin-singlet ground state. }
\label{fig3}
\end{figure}

Spectra exhibit features characteristic for the different regimes and
ZBAs of different origins emerge as the gap $\Delta$ decreases, see
Fig.~\textcolor{blue}{\ref{fig3}(a)}. In the doublet regime, an extremely narrow {\it
needle-like} Kondo resonance at $\omega=0$ coexists with Shiba
resonances at $\omega\neq 0$. The needle is due to the Kondo screening
of the magnetic doublet and has a very low Kondo temperature $T_K^N$
due to small $\Gamma_N$. In the DS cross-over region, the Shiba
resonances merge with this needle Kondo resonance to produce an {\it
enhanced ZBA} ($\Delta=0.002$, dashed line) with large height and
spectral weight. The maximum weight of this peak corresponds quite
accurately to the value of $\Delta$ where $\langle d_\uparrow
d_\downarrow \rangle$ changes sign (criterion a). Decreasing $\Delta$
further, the peak first reduces in amplitude and then splits,
signalling the end of the cross-over into the singlet phase,
characterized by two Shiba resonances at finite energy. Surprisingly,
the splitting happens precisely at the DS transition line of the
$\Gamma_N=0$ case.
%
%
%

%
%

In Fig.~\textcolor{blue}{\ref{fig3}(b)}, we plot the anomalous spectral function
$B(\omega)$ which provides information about the induced pairing in
the quantum dot. For $\Gamma_N=0$, inside the gap there would only be
delta peaks corresponding to the YSR states with positive weight for
$\omega>0$ in the doublet phase, and negative sign in the singlet
phase. For finite $\Gamma_N$, the YRS delta peaks are broadened into
resonances and the DS cross-over corresponds to a transition case
featuring both positive and negative spectral weight in $B(\omega)$
for $\omega>0$. Deeper in the doublet phase ($\Delta=0.0004$ case), we
observe an important detail: although the anomalous spectral function
has predominately positive weight for $\omega>0$, corresponding to an
overall doublet character, there is a negative low-weight peak at low
frequencies which corresponds to the needle-like ZBA (inset, indicated
by an arrow). This small peak allows to rigorously ascribe the needle
ZBA to a Kondo singlet ground state. The anomalous spectrum changes
sign at the DS point ($\Delta=0.002$, dashed). This sign change can be
identified as the point where the integrated weights
$\Omega^{\pm}\equiv\int_{0}^{\Delta} d\omega B^{\pm}(\omega)$, with
$B^{\pm}(\omega)$ being the positive and negative parts of
$B(\omega)$, are equal (criterion c). Beyond this point
($\Delta=0.001$ in the figure), $B(\omega)<0$ for $\omega>0$, as
expected for a singlet.

\subsection{$\Gamma_N$ dependence}

\begin{figure}
  \centering
  \includegraphics[angle=0,width=0.5\textwidth]{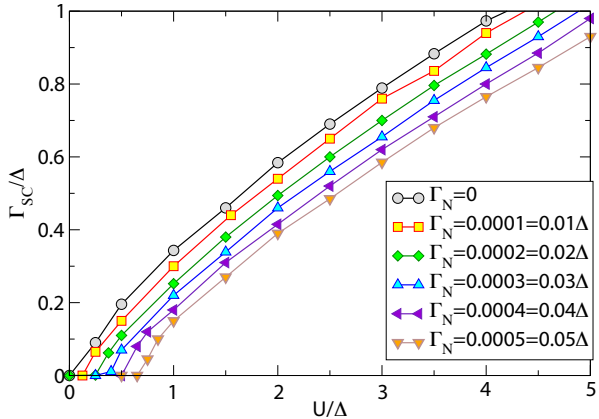}
  \caption{(Color online) Phase diagram (according to the criterion a)
  for the $DS$ cross-over when $\Gamma_{SC}$ and $U$ are tuned for
  different values of $\Gamma_{N}$. The gap is fixed to $\Delta=0.01$.
  Compared to Fig.~\textcolor{blue}{\ref{fig2}}, here $\Delta$ is fixed rather than $U$. For this
  reason, the behavior near the origin is different. In this figure,
  the origin corresponds to the non-interacting $U \to 0$ limit, while
  in Fig.~\textcolor{blue}{\ref{fig2}} the origin corresponds to the large-gap $\Delta \to
  \infty$ limit. }
    \label{fig4}
    \end{figure}

\begin{figure}
  \centering
\includegraphics[width=0.5\textwidth]{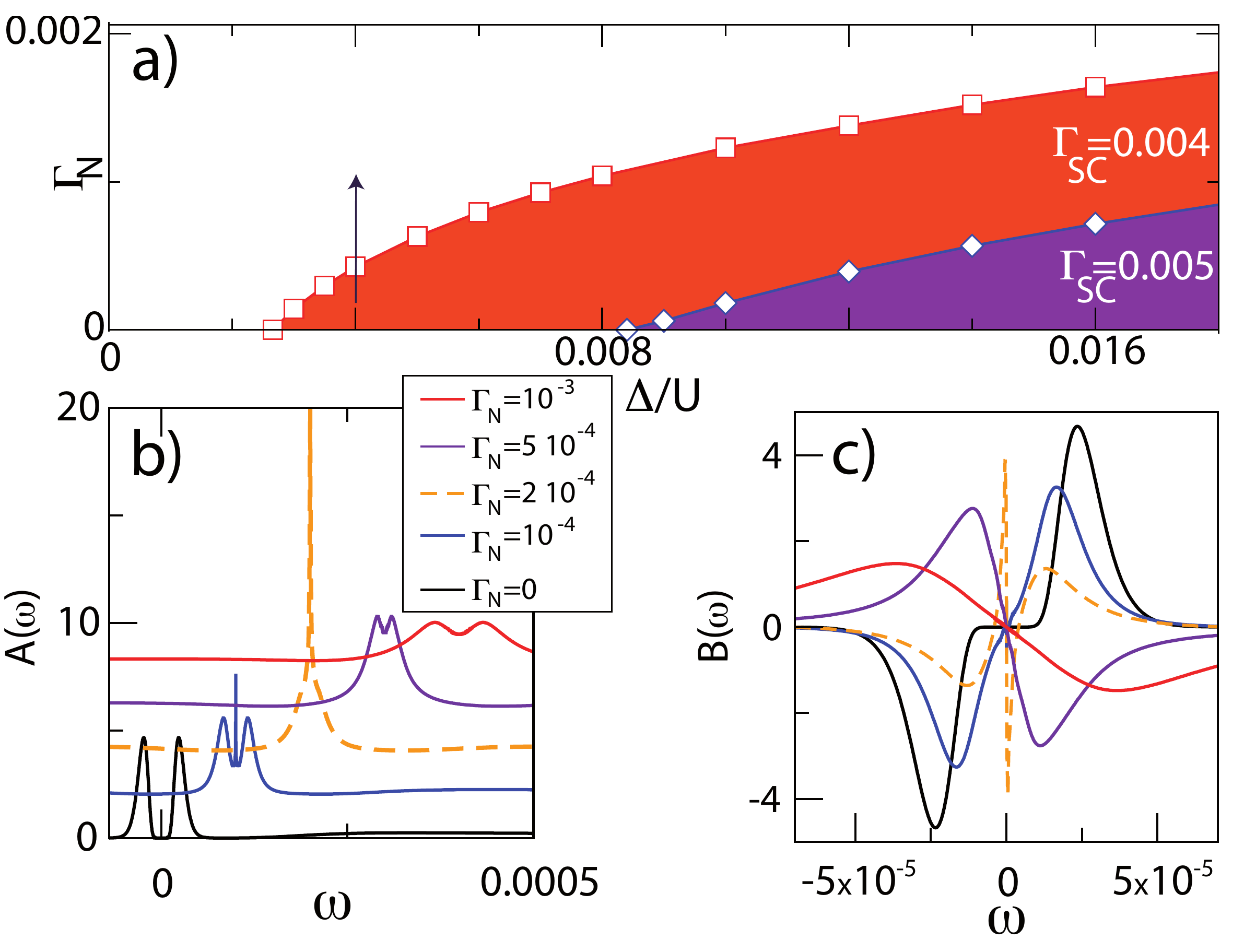}
\caption{(Color online) a) Phase diagram for fixed $U=0.05$ and
two values of $\Gamma_{SC}$ as a function of $\Gamma_N$ and $\Delta$. The colored
areas denote doublet regions. b) Spectral function (curves offset) and
c) anomalous spectrum as we increase $\Gamma_N$ along the direction
of the arrow in panel a ($\Delta=0.0002$ and $\Gamma_{SC}=0.004$).}
\label{fig5}
\end{figure}

To better understand the role of $\Gamma_N$, we summarize the results
of comprehensive calculations in Fig.~\textcolor{blue}{\ref{fig4}} where we distinguish
the two regimes when both $\Gamma_{SC}$ and $U$ are tuned at fixed
$\Delta=0.01$. Even weak coupling to the normal lead has a
considerable effect on the phase diagram, the main effect being the
significant downward shift (as a function of $\Gamma_{SC}$) of the
boundary between the singlet and doublet regimes as $\Gamma_N$
increases from zero at a fixed value of $U$. Alternatively, one may
study changes in the phase diagram as both $\Gamma_N$ and $\Delta$
vary for fixed $U$ and $\Gamma_{SC}$. These results are shown in
Fig.~\textcolor{blue}{\ref{fig5}}. Again, small values of $\Gamma_N$ (the ranges shown
on the vertical axis are always smaller than $\Gamma_{SC}$) can change
the phase diagram and induce DS transitions.

\begin{figure*}
    \centering
    \includegraphics[height=0.35\textwidth]{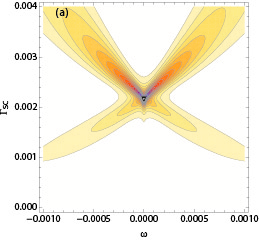}
    \includegraphics[height=0.35\textwidth]{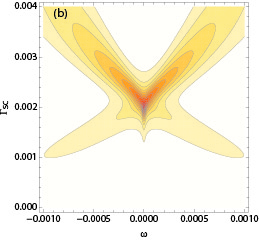}\\
    \includegraphics[height=0.35\textwidth]{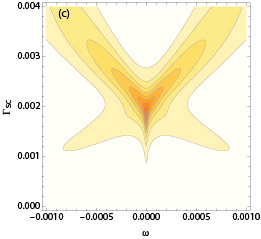}
    \includegraphics[height=0.35\textwidth]{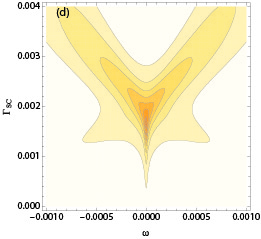}
   \caption{(Color online) Impurity spectral function $A(\omega)$ vs.
   coupling to the superconducting lead $\Gamma_{SC}$. Calculations are performed at
   fixed $U=0.01$ and $U/\Delta=5$, and plotted for a range of
   increasing coupling to the normal-state lead $\Gamma_N$: (a) $\Gamma_N$=0.0001,  (b) $\Gamma_N$=0.0002,  (c) $\Gamma_N$=0.0003 and (d) $\Gamma_N$=0.0004.
   Note the progressively wide range of $\Gamma_{SC}$ where a
   zero-bias resonance exsits as $\Gamma_N$ increases.}
	    \label{fig6}
	    \end{figure*}

The effect of $\Gamma_N$ on the width of spectral features--and
consequently on the extent of the cross-over region--is presented also
in Fig.~\textcolor{blue}{\ref{fig6}} through the $\Gamma_{SC}$ dependence of the
spectral function computed for a range of couplings to the
normal-state lead $\Gamma_N$. The plots very graphically demonstrate
the broadening effect of finite $\Gamma_N$. While in the $\Gamma_N \to
0$ limit, the crossing of the doublet and singlet states at $\omega=0$
is a discrete event that occurs at a well-defined value of
$\Gamma_{SC}$, for non-zero $\Gamma_N$ we see that there is an
extended range of $\Gamma_{SC}$ for which an observable resonance is
pinned at the Fermi level. This range corresponds to the extent of the
DS cross-over, indicated in Fig.~\textcolor{blue}{\ref{fig2}} by shading.

\subsection{Strong Coulomb interaction regime}

\begin{figure}
\centering
\includegraphics[width=0.5\textwidth]{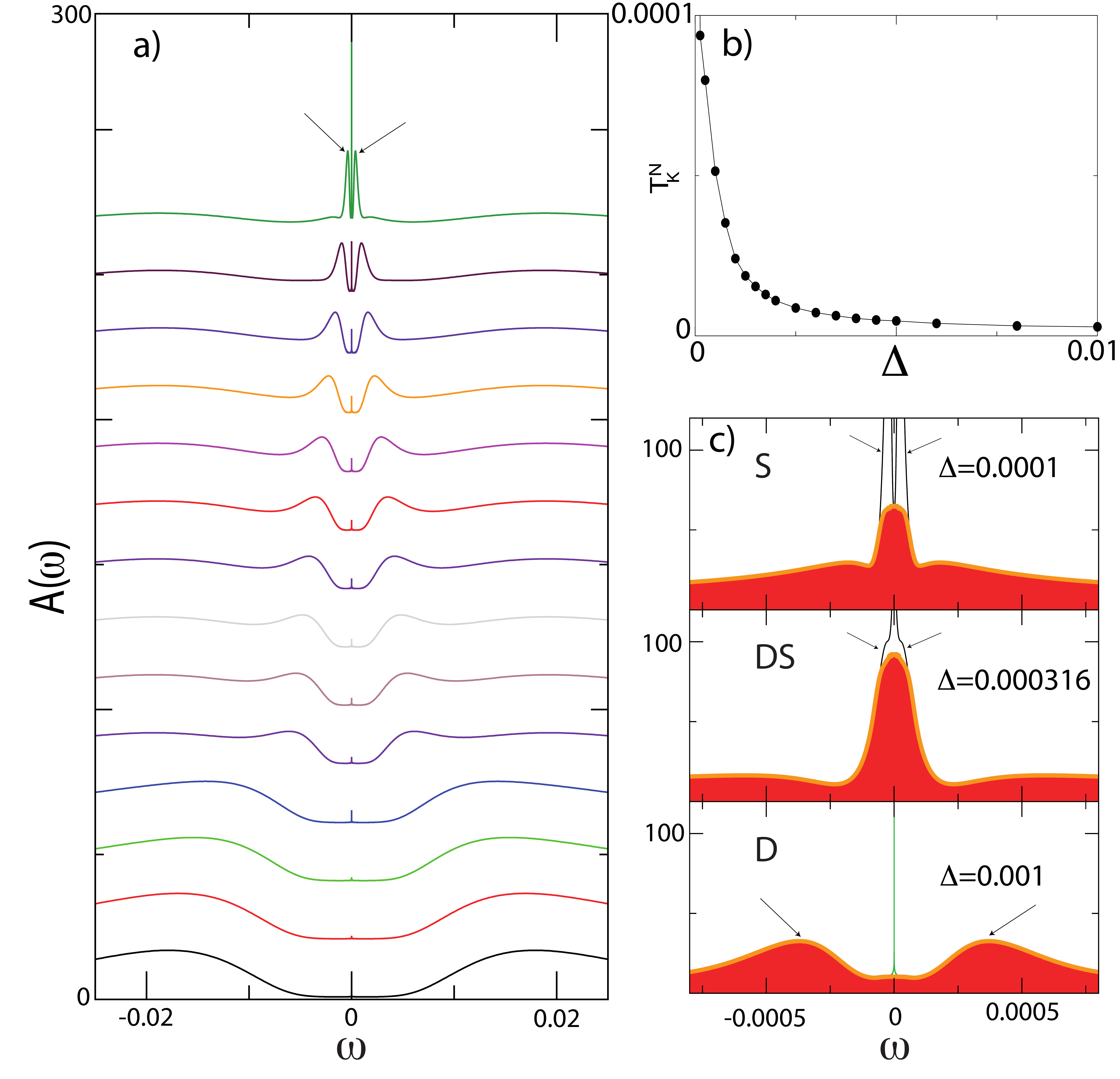}
\caption{(Color online) Large $U$ case, $U=0.05$. a) Spectral
densities for decreasing $\Delta$ (curves offset) from $\Delta=0.05$
(bottom) to $\Delta=0.001$ (top). Other parameters:
$\Gamma_{SC}=0.004$, $\Gamma_N=0.0004$. b) $T_K^N$ versus $\Delta$ for
the same parameters. c) Spectral functions at finite temperature
$T=0.002U$ (shaded curves) as the gap decreases (for a typical
charging energy of $U\sim \unit[1]{meV}$, the temperature used in the
calculations would correspond to $T\sim \unit[2]{\mu eV}\sim
\unit[23]{mK}$). The corresponding zero-temperature results are shown
as thin lines. }
 %
\label{fig7}
\end{figure}

In the strong Coulomb interaction regime with large $U/\Delta$ ratio,
one can reduce the gap to very small values before crossing over to
the singlet ground state. The phase diagram in this regime, shown in
Fig.~\textcolor{blue}{\ref{fig5}} for two fixed values of $\Gamma_{SC}$, demonstrates
the role of $\Gamma_N$: an increasing $\Gamma_N$ can drive a DS
crossover (see also panels b and c) which, for the chosen parameter
set, occurs at $\Gamma_N \approx 2\times 10^{-4}=5\times
10^{-2}\Gamma_{SC}$.
For large $U$,
the spectra are quite different from the ones shown in Fig.~\textcolor{blue}{\ref{fig2}}.
Starting from a typical configuration with a needle (Fig.~\textcolor{blue}{\ref{fig7}(a)},
bottom curves), the spectral function evolves for decreasing gap into
a characteristic shape which, apart from the needle Kondo peak, has
two large Coulomb blockade peaks, two BCS gap-edge singularities, and two emerging
Shiba satellites (top curve). Despite the significant changes in the overall
shape for varying $\Delta$, these spectra all belong to the doublet regime.

\section{Role of finite temperatures} 

The role of finite $T$ is most pronounced in the doublet regime. The
Kondo temperature of the needle peak, $T_K^N$, depends exponentially
on $\Gamma_N$, but not in the standard way since $U$ is renormalized
by the screening from the superconducting lead (see Appendix C).
Importantly, $T_K^N$ grows as $\Delta$ decreases, as indicated by the
numerical results in Fig.~\textcolor{blue}{\ref{fig7}(b)} and by the Schrieffer-Wolff
transformation which shows an enhanced Kondo exchange coupling as
$\Delta$ is reduced, as demonstrated in Appendix B. 
In the large-$U$ regime, this temperature scale may be of the order or
larger than the splitting of YRS states after the DS transition. This
results in large ZBAs as the gap closes, see Fig.~\textcolor{blue}{\ref{fig7}(c)}. Similar
features in the spectrum could be attributed to emergent MBS
\textcolor{blue}{\cite{Das12ZBA,Deng12,Finck:PRL13}}. Therefore, a word of caution about
this interpretation is in order.

\section{Role of magnetic fields}

\subsection{Field-induced zero-bias anomaly}

\begin{figure}
 \centering
\includegraphics[width=0.5\textwidth]{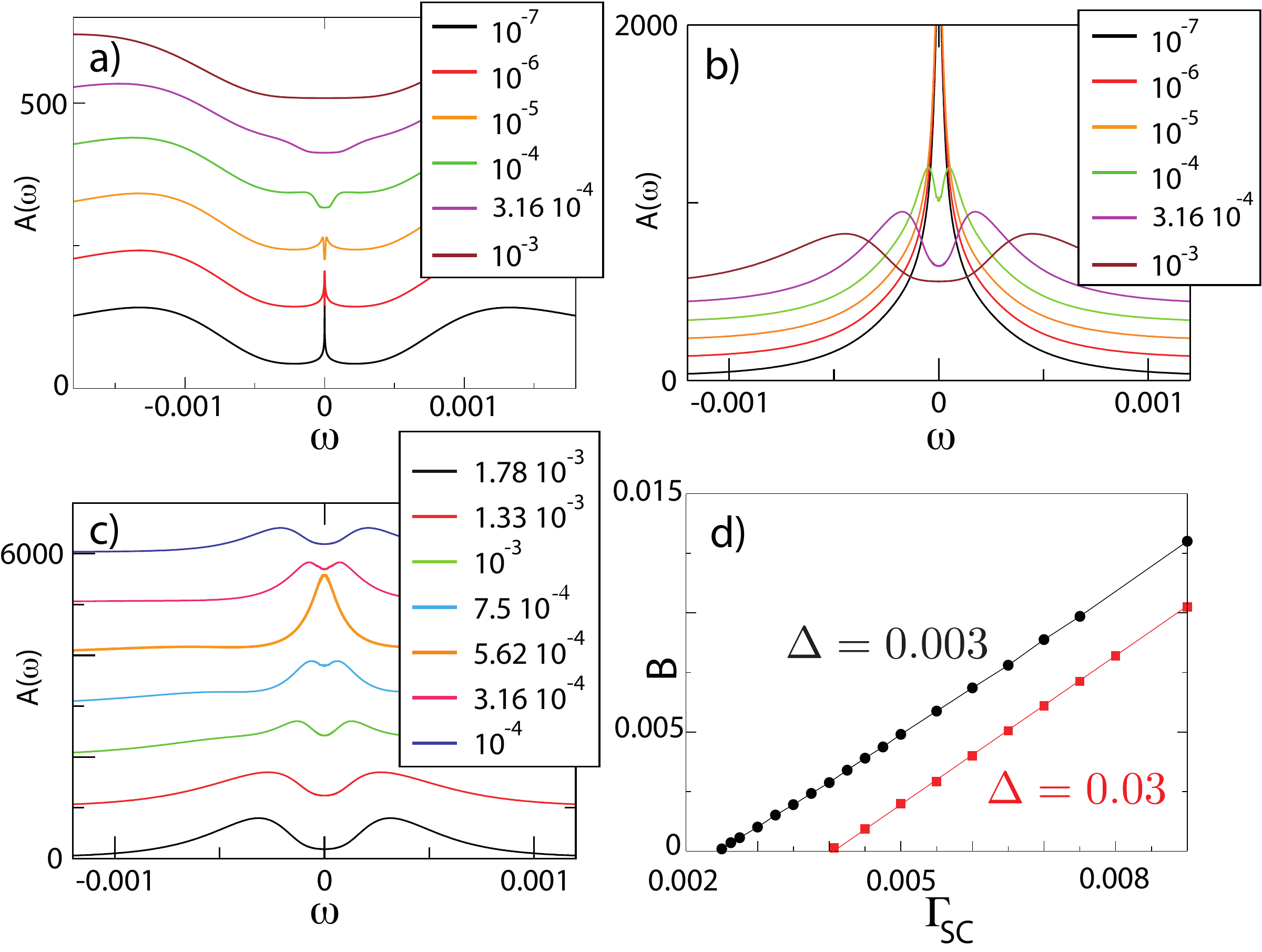}
\caption{(Color online) Effect of the magnetic field on the spectral
functions. We plot the spin-averaged spectral function for a range of
magnetic fields $B$ in a) doublet ($\Gamma_{SC}=0.001$), b) DS
crossover ($\Gamma_{SC}=0.00225$), and c) singlet 
($\Gamma_{SC}=0.003$) regimes (curves vertically offset for clarity).
Other parameters are $U=0.01$, $\Gamma_N=0.0002$, $\Delta=0.003$. d)
Position of the parity crossing in magnetic field versus
$\Gamma_{SC}$.
}
\label{fig8}
\end{figure}

Magnetic field is used to induce topologically nontrivial phases with
Majorana states in nanowires; hence it is interesting to see whether
ZBAs can be generated by the field also in the quantum dot system. 
The spectra for a range of fields are presented in Fig.~\textcolor{blue}{\ref{fig8}}. In
the doublet regime (panel a), we observe outward shift of the Shiba
states induced by enlarged DS excitation energy as $B$ is increased,
as well as the Zeeman splitting of the needle ZBA leading to a
pronounced dip structure at moderate $B$. In the DS cross-over regime
(panel b) where the Kondo peak is already merged with Shiba states, we
see the splitting of this collective ZBA. The most interesting case is
the S regime (panel c), where parity crossings occur as one of the
Zeeman split doublet states becomes the new ground state at some
finite $B$: at this point a sizeable ZBA is formed, in agreement with
the experiments of Ref.~\textcolor{blue}{[\onlinecite{Lee14}]}. We note that the combined
action of the above phenomenology with the previously discussed DS
transitions as one reduces the gap would lead to ZBAs that split and
re-form, similar to the observations in e.g.
Ref.~\textcolor{blue}{[\onlinecite{Finck:PRL13}]}.

\begin{figure*}
    \centering
        \includegraphics[height=0.35\textwidth]{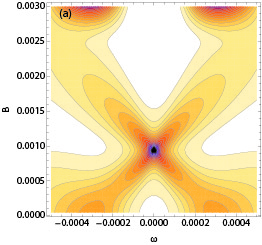}
    \includegraphics[height=0.35\textwidth]{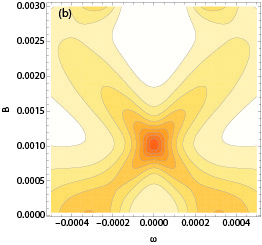}\\
    \includegraphics[height=0.35\textwidth]{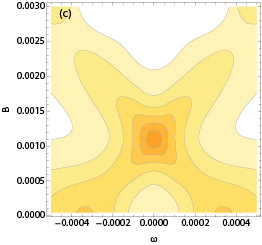}
    \includegraphics[height=0.35\textwidth]{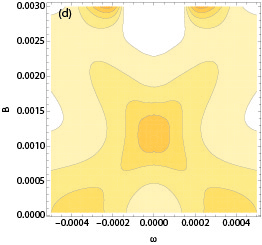}
   \caption{(Color online) Impurity spectral function $A(\omega)$ as a
   function of the 
   magnetic field $B$. The calculations are performed at fixed $U=0.01$,
   $\Gamma_{SC}=0.003$, $\Delta/U=0.3$, and plotted for a range of
   increasing coupling to the normal-state lead $\Gamma_N$: (a) $\Gamma_N$=0.0001,  (b) $\Gamma_N$=0.0002,  (c) $\Gamma_N$=0.0003 and (d) $\Gamma_N$=0.0004.
   Note that $\Gamma_N \ll \Gamma_{SC}$ for all cases considered.}
	    \label{fig9}
	    \end{figure*}

In Fig.~\textcolor{blue}{\ref{fig9}} we plot the dependence of the spectral function
on the external magnetic field for a range of hybridization strengths
to the normal-state lead, $\Gamma_N$. For small $\Gamma_N$, the
crossing of the lower doublet state with the single YSR state is
characterized by a very pronounced zero-bias anomaly occuring at a
well defined value of the magnetic field. As $\Gamma_N$ increases, the
spectral features become more diffuse, thus there is an extended range
of magnetic fields with enhanced spectral densities near the Fermi
level. This is similar to the behavior observed in some experiments
aiming at the detection of Majorana bound states.

\subsection{Linear $B$ vs. $\Gamma_{SC}$ dependence}

The Zeeman-induced ZBA in the singlet regime is continuously connected
with the DS crossing at $B=0$ for a different value of $\Gamma_{SC}$.
In fact, our numerical results show that the position in B field of
this ZBA depends \emph{linearly} on $\Gamma_{SC}$ for any value of
$\Delta$ (Fig.~\textcolor{blue}{\ref{fig8}(d)}). This is highly surprising, since the
singlet-doublet splitting is non-linear in $\Gamma_{SC}$, and the
Zeeman splitting is non-linear in both $\Gamma_{SC}$ and $B$;
nevertheless, the intersection happens along a straight line in the
$(B,\Gamma_{SC})$ plane as long as the system is particle-hole
symmetric. 

This linear dependence can be obtained analytically in the small
$\Gamma_N$ limit by studying the conditions for the occurence of the
subgap states exactly at the Fermi level at $\omega=0$. We will thus
focus on the $|\omega| \ll \Delta$ limit, noting that this is not at
all the same as the $\Delta \to \infty$ limit. We assume that the magnetic
field is applied along the $z$ axis.

The interaction effects are fully described
by the self-energy matrix, introduced through the Dyson equation
\begin{equation}
\hat{G}(z)^{-1}=\hat{G}^{(0)}(z)^{-1}
-\hat{\Sigma}(z),
\end{equation}
where the non-interacting Green's function matrix is
\begin{equation}
\label{e1}
\hat{G}^{(0)}(z)^{-1}=
z - \epsilon \tau_3 - E_Z \sigma_3 - V^2 \tau_3 \frac{1}{N}
\sum_k g_k(z) \tau_3.
\end{equation}
Here $z$ is the complex frequency argument (taken to be
$z=\omega+i\delta$ at the end of the calculation to obtain the
retarded Green's functions), $E_Z=g\mu_B B/2$ is the Zeeman energy,
$V$ is the coupling to the superconducting
lead (the normal lead is not considered in this section), $N$ is the
number of $k$ states in the lead, $g_k(z)$ is the Green's function for
an electron in the superconducting lead and, finally, $\tau_i$ are
Pauli matrices in the Nambu (particle-hole) space, while $\sigma_i$
are Pauli matrices in the spin space. For magnetic field applied along
the $z$ axis, it is possible to work either with the $2 \times 2$
Nambu structure with $\Psi=(d_\uparrow d_\downarrow^\dag)^T$, or with
the $4 \times 4$ Nambu structure with $\Psi=(d_\uparrow d_\downarrow
d_\downarrow^\dag d_\uparrow^\dag)^T$. In the latter case, the
$2\times 2$ submatrices are actually diagonal. In the former case, the
$\sigma_3$ matrix in Eq.~\textcolor{blue}{\eqref{e1}} needs to be replaced by the
identity.

Since
\begin{equation}
g_k(z)^{-1}=z-\epsilon_k\tau_3+\Delta \tau_1\sigma_3,
\end{equation}
one finds
\begin{equation}
g_k(z)=\frac{z+\epsilon_k\tau_3-\Delta
\tau_1\sigma_3}{z^2-(\epsilon_k^2+\Delta^2)}.
\end{equation}
Summing over $k$ in Eq.~\textcolor{blue}{\eqref{e1}}, one obtains
\begin{equation}
G^{(0)}(z)^{-1} = z-\epsilon\tau_3-E_Z \sigma_3+\Gamma_{SC}
\frac{(z+\Delta \tau_1\sigma_3)}{E(z)},
\end{equation}
where the last term is the self-energy originating from the coupling
with the superconducting lead. $E(z)$ can be analytically continued to
$E(\omega)=\sqrt{\Delta^2-\omega^2}$. In the $\omega \to 0$ limit,
$E(0)=\Delta$ and the coupling self-energy reduces to
$\Gamma_{SC}\tau_1\sigma_3$. Note that in this limit the gap $\Delta$
disappears from the problem such that $\Gamma_{SC}$ plays the role of
an effective pairing term. 

The Shiba states are identified as the poles of the Green's function
inside the gap:
\begin{equation}
\det[G^{-1}(z)]=0,
\end{equation}
where $z$ needs to be on the real axis for a true bound state, while
resonances correspond to true solutions with a small imaginary component
(this would be the case for $\Gamma_N \neq 0$).
\begin{figure}
  \centering
  \includegraphics[width=0.45\textwidth]{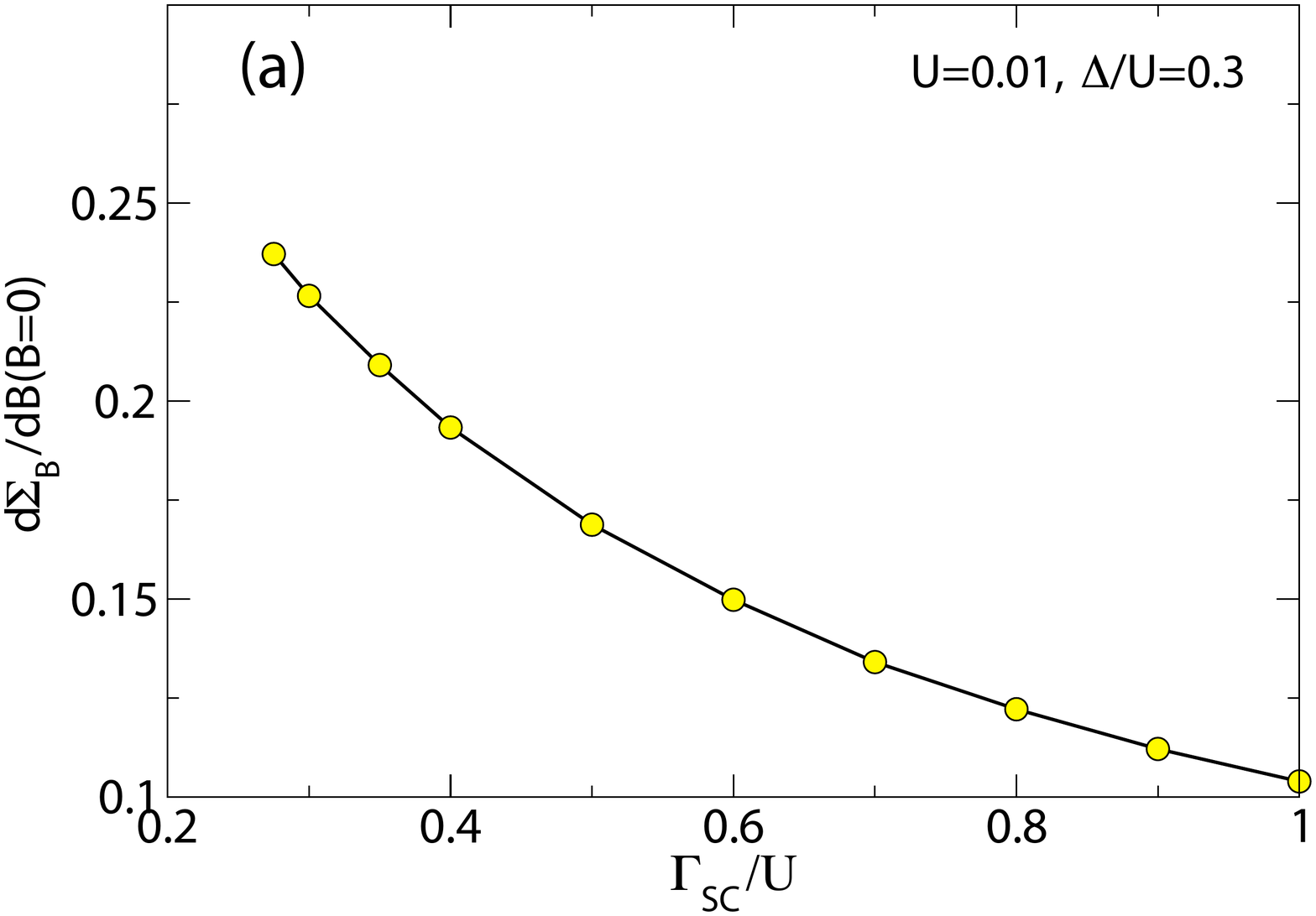}
  	  \includegraphics[width=0.45\textwidth]{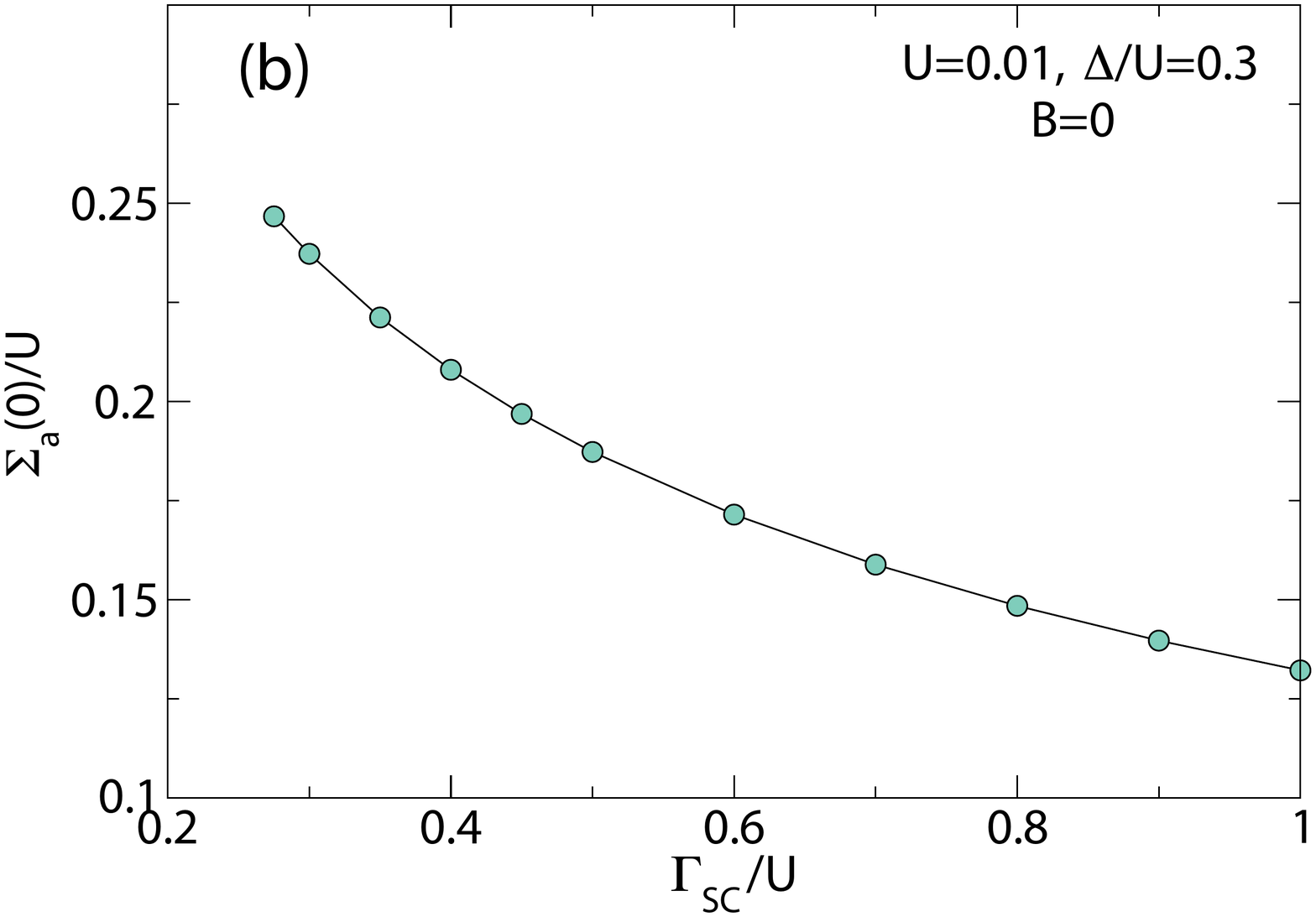}
    \caption{(Color online) (a) Slope of the real part of $\Sigma_B(B)$
      self-energy function. This quantity can be interpreted as the
        renormalization of the effective $g$-factor due to interactions.  (b) Zero-frequency value of the real
	      part of the anomalous self-energy, $\mathrm{Re}\Sigma_a(\omega=0)$
	        in the singlet regime, $\Gamma>\Gamma_{DS}$.
	}
	\label{fig10}
	\end{figure}
In the absence of interactions, the condition for a sub-gap state
takes the following form:
\begin{equation}
z^2-\epsilon^2-E_Z z + E_Z^2 - \Gamma_{SC}^2 \frac{\Delta^2-z^2}{E(z)^2}
+\frac{2z(z-E_Z)\Gamma_{SC}}{E(z)}=0.
\end{equation}
Taking the $|z| \ll \Delta$ limit, this yields
\begin{equation}
\label{e2}
E_Z^2= \Gamma_{SC}^2 + \epsilon^2.
\end{equation}
Interestingly, this condition for a Zeeman-induced zero-energy YSR
state in a non-interacting quantum dot is the same as the one in
Eq.~\textcolor{blue}{\eqref{criterion}} for obtaining the MBS in a nanowire (as we
mentioned, in the $z\rightarrow 0$ limit $ \Gamma_{SC}$ plays the role
of an effective pairing term $\Delta_*$, while $\epsilon$ plays the
role of a chemical potential
in the quantum dot). 

Eq.~\textcolor{blue}{\eqref{e2}} can be easily generalized to the interacting case. The
structure of the self-energy matrix is
\begin{equation}
\hat{\Sigma}(z)=\begin{pmatrix}
\Sigma_\uparrow(z) & 0 & \Sigma_a(z) & 0 \\
0 & \Sigma_\downarrow(z) & 0 & -\Sigma_a(-z) \\
\Sigma_a(z) & 0 & -\Sigma_\downarrow(-z) & 0 \\
0 & -\Sigma_a(-z) & 0 & -\Sigma_\uparrow(-z) 
\end{pmatrix},
\end{equation}
where $\Sigma_\sigma(z)$ are the regular self-energy components,
while $\Sigma_a(z)$ is the anomalous component. To study the positions
of the sub-gap peaks, a low-order expansion can be performed:
\begin{equation}
\hat{\Sigma}(z) = \hat{\Sigma}(0) + \hat{\Sigma}'(0) z =
\hat{\Sigma}(0) + (1-\hat{Z}^{-1})z,
\end{equation}
where $\hat{Z}$ is the (matrix-valued) quasiparticle renormalization
factor $\hat{Z}^{-1}=1-\hat{\Sigma}'(0)$ whose deviation from the
identity matrix quantifies the strength of the interaction effects. In
fact, for our consideration of the zero-crossing, we truncate the
expansion at the first term. This is an important observation
which holds in general: {\sl the condition for the zero-energy
Shiba state does not depend explicitly on the quasiparticle
renormalization factor (i.e., on the Kondo temperature)}.
We are thus only interested in the
zero-frequency values, $\hat{\Sigma}(0)$. These are purely real, since
the self-energy has zero imaginary part inside the superconducting
gap. 
%
%
We insert the self-energy matrix in Eq.~\textcolor{blue}{\eqref{e1}}, evaluate the
determinant in the $|\omega| \ll \Delta$ limit, and after some lengthy
algebra obtain the following expression:
\begin{equation}
(E_Z+\Sigma_B)^2 = (\Gamma_{SC}-\Sigma_a)^2 + (\epsilon+\Sigma_0)^2,
\end{equation}
where we have introduced the spin-averaged normal self-energy
$\Sigma_0\equiv\frac{1}{2}(\Sigma_\uparrow(0)+\Sigma_\downarrow(0))$
and the spin component
$\Sigma_B\equiv\frac{1}{2}(\Sigma_\uparrow(0)-\Sigma_\downarrow(0))$,
with $\Sigma_\sigma(0)=Un_{\bar{\sigma}}$.
This equation maintains the structure of Eq.~\textcolor{blue}{\eqref{e2}}, the only
new effects are the interaction-induced shifts. In the particle-hole
symmetric case, one has $\Sigma_0=U/2$ and $\epsilon=-U/2$, thus the
last term drops out. Then
\begin{equation}
E_Z+\Sigma_B = \Gamma_{SC}-\Sigma_a.
\end{equation}
This equation turns out to describe a linear relation between $E_Z$
(i.e., field $B$) and $\Gamma_{SC}$ despite the non-trivial
$\Gamma_{SC}$-dependence of the self-energies $\Sigma_B$ and
$\Sigma_a$, since $\Sigma_B$ is proportional to $B$ to a very good
approximation, $\Sigma_B=c(\Gamma_{SC})B$, and there appears to be a
connection between the Fermi-level derivative of the spin-dependent
self-energy
$c(\Gamma_{SC})=\mathrm{d}\Sigma_B/\mathrm{d}B|_{\omega=0}$ and the
anomalous self-energy $\Sigma_a(\Gamma_{SC})$, see Figs.~\textcolor{blue}{\ref{fig10}(a)}
and \textcolor{blue}{\ref{fig10}(b)}. Plotting
$(\Gamma-\Sigma_a(\Gamma))/(1+\mathrm{d}\Sigma_B/\mathrm{d}B)$ as a
function of $\Gamma$, one obtains a straight line with a slope close
to 2.
	        \begin{figure}
         \centering
           \includegraphics[width=0.32\textwidth]{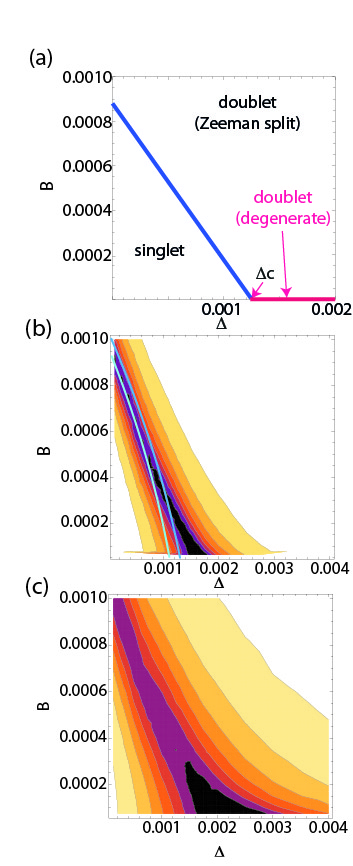}
\caption{(Color online) (a) Phase diagram in the $(B,\Delta)$
	plane for $\Gamma_N=0$. At $\Delta_c\sim 0.0012$, the ground state of the system at $B=0$ changes from singlet to doublet. (b,c) Zero-frequency spectral function
	$A(\omega=0)$ plotted as a function of the gap $\Delta$ and
	the external magnetic field $B$, revealing the behavior of the
	zero-bias anomaly in the $(\Delta,B)$ plane. The coupling to
	the normal-state lead is (b) $\Gamma_N=0.0002$, and (c)
	$\Gamma_N=0.0004$. In (b) we also plot (in blue) two possible
	lines for the evolution of the gap for increasing magnetic fields. We use the function $\Delta(B)\sim\Delta[1-0.32 B-0.1 B^2]$, based on a fitting of the experimental data from Ref.~\textcolor{blue}{[\onlinecite{Lee14}]}. Both curves correspond to gap values $\Delta=0.0011$ and $0.0013$, respectively, which are located on either side of the $\Gamma_N=0$ transition around $\Delta_c$. Rest of parameters: $\Gamma_{SC}=0.002$ and $U=0.01$.}
           \label{fig11}
           \end{figure}
		
		We also note that for zero-field, the DS cross-over is defined
		through
		\begin{equation}
		\Gamma_{SC}=\Sigma_a(\Gamma_{SC}).
		\end{equation}

We conclude that the ZBA occurs for
%
%
\begin{equation}
\label{eq2}
\tilde{E}_Z^2 = \tilde{\Gamma}_{SC}^2 + \tilde{\epsilon}^2.
\end{equation}
Here, tilde quantities represent parameters renormalized by
interactions $X\rightarrow \tilde{X}\equiv X+Re\Sigma(\omega=0)$. At
the particle-hole symmetric point the last term drops out so that
\begin{equation}
\pm\tilde{E}_Z=\tilde{\Gamma}_{SC}.
\end{equation}
We stress again that this linear relation for arbitrary $U$ and
$\Delta$ is remarkable since the corresponding self-energies
renormalizing the bare parameters, like for instance the renormalized
g-factor that can be extracted from $\tilde{E}_Z$, are themselves
non-linear functions of $\Gamma_{SC}$.

Interestingly, Eq.~\textcolor{blue}{\eqref{eq2}} still has the same structure as Eq.~\textcolor{blue}{\eqref{e2}}.
Therefore, the general condition for Zeeman-induced parity crossings of YSR bound states, fully taking into account interactions, and the condition for reaching a topological phase in a
non-interacting nanowire (Eq.~\textcolor{blue}{\eqref{criterion}}) are still analogous. 

\subsection{Zero-bias anomalies studied in the $(\Delta,B)$ plane}
In experiments performed on nanowires exposed to external magnetic field, the
role of the field is two-fold: (a) it leads to Zeeman
splitting of the doublet YSR states, and (b) it suppresses the BCS
pairing parameter $\Delta$. Up to now, we have presented results
computed for varying $B$ at fixed $\Delta$. For completeness, we now
provide some results computed as a function of both $B$ and $\Delta$:
the actual experimental situation corresponds to some
$\Delta=\Delta(B)$ curve in this plane.

			     
In Fig.~\textcolor{blue}{\ref{fig11}(a)} we present the phase diagram in the $\Gamma_N
\to 0$ limit. 
For small $\Delta$,
the ground state at zero field is a singlet. As $B$ increases, one of
the Zeeman-split doublet levels is brought down in energy and
eventually becomes the new ground state. In this part of the diagram,
we observe linear dependence between $\Delta$ and $B$ at the
doublet-singlet transition. Note that this is yet another unexpected
linearity, different (but related) to the one in the
$(\Gamma_{SC},B)$ plane discussed above. 

The effect of the coupling to the normal-state leads is demonstrated
in Fig.~\textcolor{blue}{\ref{fig11}(b,c)}, where we plot the dependence of the spectral
function at zero frequency, $A(0)$, on $\Delta$ and $B$. The spectra
are strongly enhanced (i.e., feature a zero-bias anomaly) in two
regions: (i) for small $\Delta < \Delta_c \approx 0.0012$ for magnetic
fields where the singlet and doublet states cross at $\omega=0$, and
(ii) for large $\Delta > \Delta_c$ near zero-field, due to the
needle-like Kondo resonance induced directly by the normal-state
tunneling probe. We note that in this case non-zero $\Gamma_N$
strongly suppresses the linearity of the ZBA in region (i).

The precise $\Delta(B)$ function form depends on the experimental
details. To indicate the possible behavior, we overlayed two curves on
Fig.~\textcolor{blue}{\ref{fig11}(b,c)}. Both curves, based on realistic
$\Delta(B)$ dependence  for the particular experiment
described in Ref.~\textcolor{blue}{\onlinecite{Lee14}}, indicate that persisting ZBAs can be found in
this parameter plane. Both curves correspond to gap values, at $B=0$, around $\Delta_c$. The lighter curve corresponds to the case where
upon increasing $B$, the ZBA appears and persists practically until
the gap closure. The darker curve corresponds to the case where the
ZBA first appears and then splits again before the gap is ultimately
closed.

\section{Conclusion} 

We have calculated the phase diagram of an Anderson impurity in
contact with superconducting and normal-state leads by means of the
numerical renormalization group, and established that even a very weak
coupling to the normal lead perturbs the system. Our results, valid
for an arbitrary ratio $\Delta\over T_K$, are analyzed in the context
of experimental scenarios such as zero-bias anomalies induced by
parity crossing transitions of Yu-Shiba-Rusinov bound states and novel
Kondo features induced by the normal lead. In particular, we have discussed how spectral functions at
finite temperatures and magnetic fields, which can be directly linked
to experimental tunneling transport characteristics, can show
zero-energy anomalies irrespective of whether the system
is in the doublet or singlet regime. These results indicate that due caution is
needed in interpreting experiments aiming to detect Majorana bound
states since in hybrid systems Kondo physics and parity crossings may
manifest in unanticipated ways. 

We have also derived the analytical
condition for the occurrence of Zeeman-induced fermion-parity switches in the
presence of interactions,  Eq.~\textcolor{blue}{\eqref{eq2}}, which bears unexpected similarities with the
condition for emergent Majorana bound states in nanowires, Eq.~\textcolor{blue}{\eqref{criterion}}. This result suggests that the physics of Zeeman-induced parity-crossings in the minimal Anderson model in contact with a superconductor is connected with the condition for emergent Majorana bound states. This similarity thus leads to an
interesting question: Is this equivalence between Eq.~\textcolor{blue}{\eqref{criterion}} and
Eq.~\textcolor{blue}{\eqref{eq2}} general? While we do not have a final answer for this, 
we note that the analogy persists for finite
spin-orbit coupling in the non-interacting regime: it has been shown
\textcolor{blue}{\cite{Stanescu:PRB13}}  that Zeeman-induced parity crossings in short
non-interacting nanowires (with finite spin-orbit coupling) smoothly
evolve towards true topological transitions as the wire becomes
longer. Whether our interacting results are also smoothly
connected with MBS physics in the finite spin-orbit case and beyond
the single quantum impurity limit remains an open question worth to be
investigated.
\begin{acknowledgments}
We thank Jens Paaske for his comments on the manuscript. Work supported by MINECO Grants No. FIS2011-23526, FIS2012-33521 and  by the
Kavli Institute for Theoretical Physics through NSF grant PHY11-25915.
R.\v Z. acknowledges the support of the Slovenian Research Agency
(ARRS) under Program P1-0044.
\end{acknowledgments}

\appendix

\section{Doublet-singlet transition induced by the normal-state lead}

\begin{figure}
  \centering
  \includegraphics[width=0.45\textwidth]{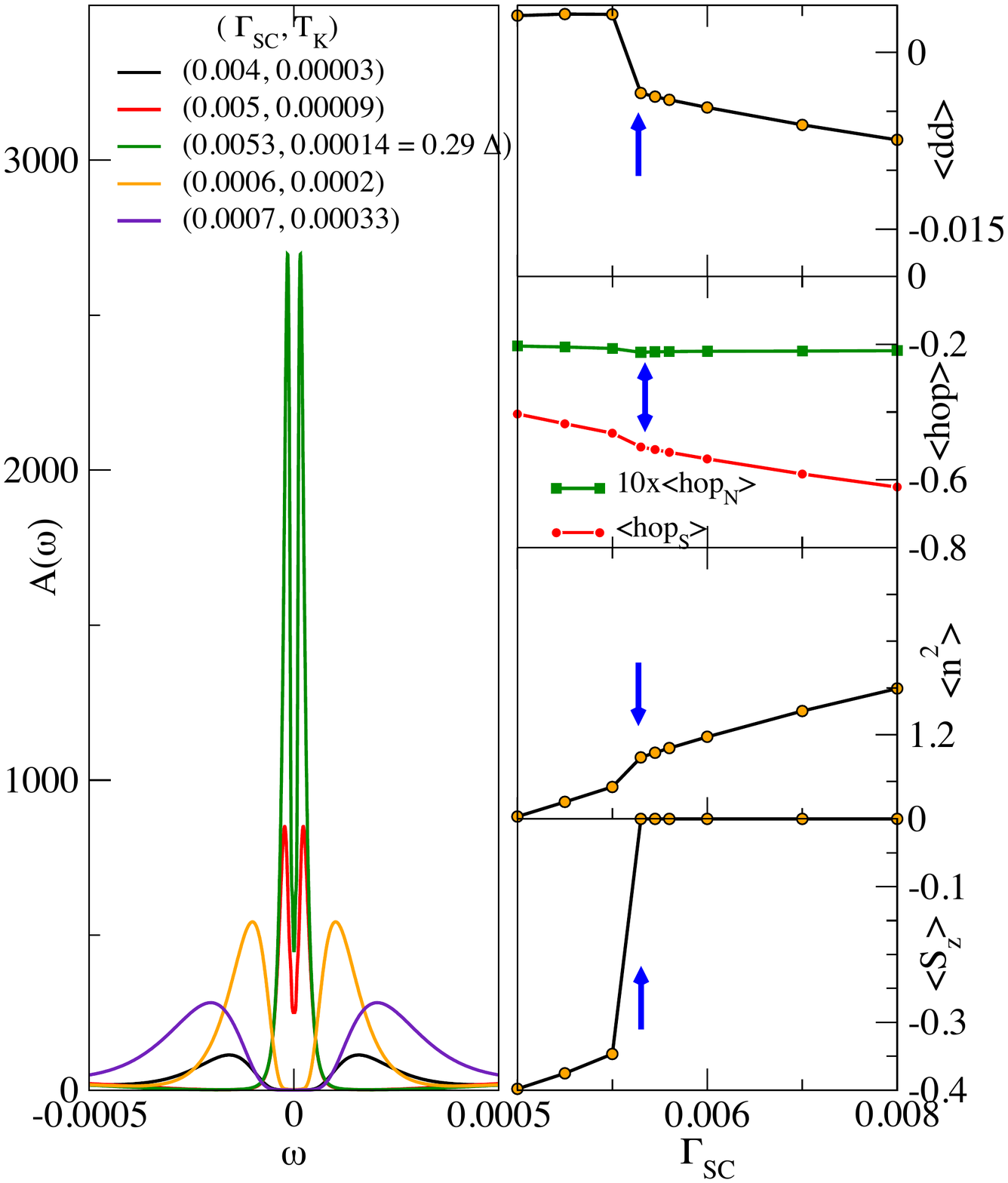}
    \caption{(Color online) Left panel: Impurity spectral functions
    for a range of hybridization
    strengths to the superconducting lead ($U=0.05$ and
    $\Delta/U=0.01$). Normal lead is nearly decoupled.
    The doublet-singlet transition occurs for $\Delta = 3.6T_K$ or
    $T_K = 0.3\Delta$, where $T_K$ is Wilson's Kondo temperature. Right
    panel: Expectation values as a function of $\Gamma_{SC}$. 
    All the criteria show a DS transition at  $\Gamma_{SC}\approx
    5.3\times 10^{-3}$ (arrows).}
    \label{figsup1}
    \end{figure}

To better illustrate how the doublet-singlet (DS) transition occurs,
we consider a situation in which the superconducting coupling
$\Gamma_{SC}$ increases while the normal-lead coupling is fixed to a
very small value $\Gamma_N=10^{-5}$ (effectively zero). Due to the
smallness of $\Gamma_N$, this situation can be identified with an
effective SC-QD setup. We plot in Fig.~\textcolor{blue}{\ref{figsup1}(a)} the impurity
spectral function when $U=0.05$, at a fixed superconductig gap value
$\Delta/U=0.01$. In order to clearly identify the doublet regions, we
include small but finite temperature and magnetic field. Finite $B$
field leads to a sizeable non-zero magnetization sufficiently deep in
the doublet phase because the magnetic moment remains unscreend; the
magnetization starts to increase at the DS transition. Also, because
$T$ is finite, one may indeed characterize the small-$\Gamma_{SC}$
phase as the doublet phase (in the zero-temperature limit, the ground
state is strictly speaking a singlet for any non-zero $\Gamma_N$). The
impurity spectral function shows the DS transition when
$\Gamma_{SC}\approx 5.3\times 10^{-3}$, which, as expected,
corresponds to $T_K\approx 0.3\Delta$.

More rigorously, one may locate the DS transition point by employing
several criteria based on the behavior of: \textit{(i)} the pairing
term $\langle d_\uparrow d_\downarrow \rangle$, \textit{(ii)} the
hopping functions $h_\alpha=\sum_\sigma \langle d_\sigma^\dag
f_{0\sigma\alpha} + \text{H.c.} \rangle$, where $f_{0\sigma\alpha}$ is
the combination of the conduction band orbitals to which the impurity
couples, \textit{(iii)} charge
fluctuations $\langle n^2\rangle$ (with $n=\sum_\sigma
d_\sigma^\dagger d_\sigma$ as the total impurity occupation), and
finally \textit{(iv)} $S_z=1/2(n_\uparrow-n_\downarrow)$ (the $z$
component of the impurity spin, i.e., the magnetization).  All these
quantities are displayed in Fig.~\textcolor{blue}{\ref{figsup1}(b)} and show
a transition at $\Gamma_{SC}\approx
5.3\times 10^{-3}$ (arrows), where all these quantities are
discontinuous. In particular,  $\langle d_\uparrow d_\downarrow
\rangle$ changes sign, while $\langle S_z \rangle$ becomes large
in the doublet phase (being essentially zero in the single phase)
due to the weak but non-zero external magnetic field.

    \begin{figure}
      \centering
      \includegraphics[width=0.5\textwidth]{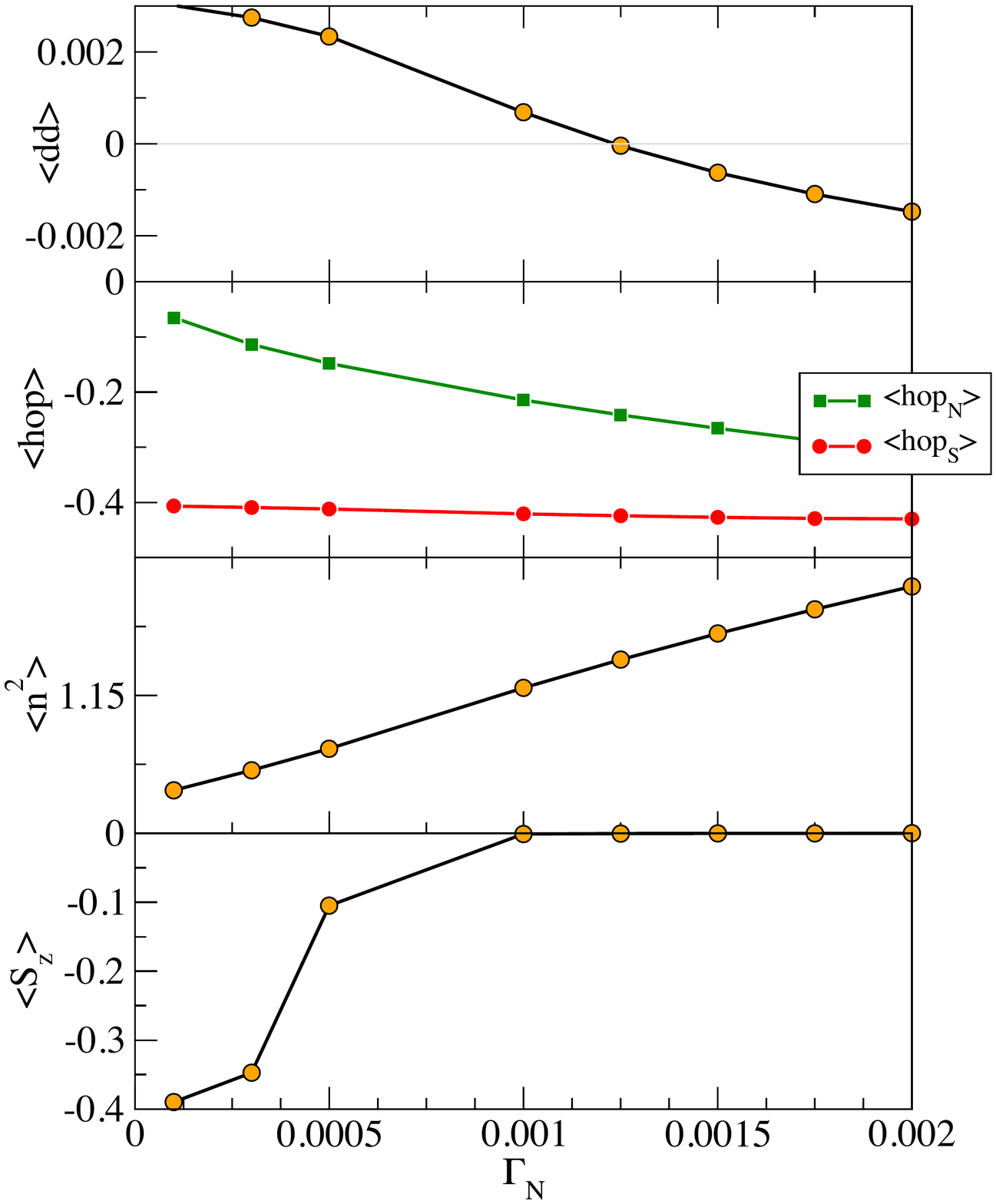}
        \caption{(Color online) 
	  Expectation values as a function of $\Gamma_N$ for a fixed
	  $\Gamma_{SC}=0.004$ (the rest of parameters are the same as in
	  Fig.~\textcolor{blue}{\ref{figsup1}}). The pairing
	  term (top panel) changes sign at $\Gamma_N
	  \approx 1.25\times 10^{-3}$ whereas the magnetization
	  (bottom panel) is non-zero for a slightly smaller value
	  $\Gamma_N=10^{-3}$.}
	  \label{figsup2}
	  \end{figure}

Now that we have established clear criteria for the DS transition, we
study how the above quantities vary as we increase $\Gamma_N$ for a
fixed $\Gamma_{SC}=0.004$ (Fig.~\textcolor{blue}{\ref{figsup2}}). As argued above,
different criteria define different values of $\Gamma_N$ at which the
system crosses over from doublet to singlet regime. Here, for
instance, the pairing term (top panel) changes sign at
$\Gamma_N=1.25\times 10^{-3}$ while the magnetization (bottom panel)
is non-zero already at $\Gamma_N=10^{-3}$. These different values of
$\Gamma_N$ according to the different criteria define a sizable
cross-over region in the phase diagram.

\begin{figure}
  \centering
  \includegraphics[angle=-90,width=0.5\textwidth]{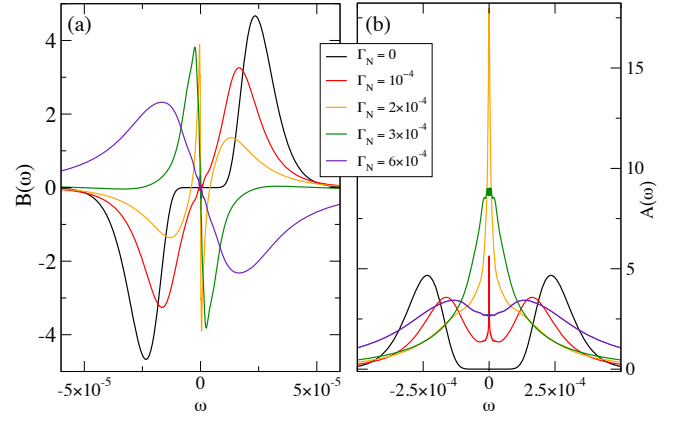}
    \caption{(Color online). (a) Characterization of the singlet phase
      by the anomalous spectral function $B(\omega)$ for a range of
        $\Gamma_N$. (b) Spectral function $A(\omega)$.
	  The model parameters are $U=0.05$, $\Gamma_{SC}=0.004$,
	  $\Delta=0.0002$.}
	  \label{figsup3}
	  \end{figure}

	  \begin{figure}
	    \centering
	    \includegraphics[width=0.5\textwidth]{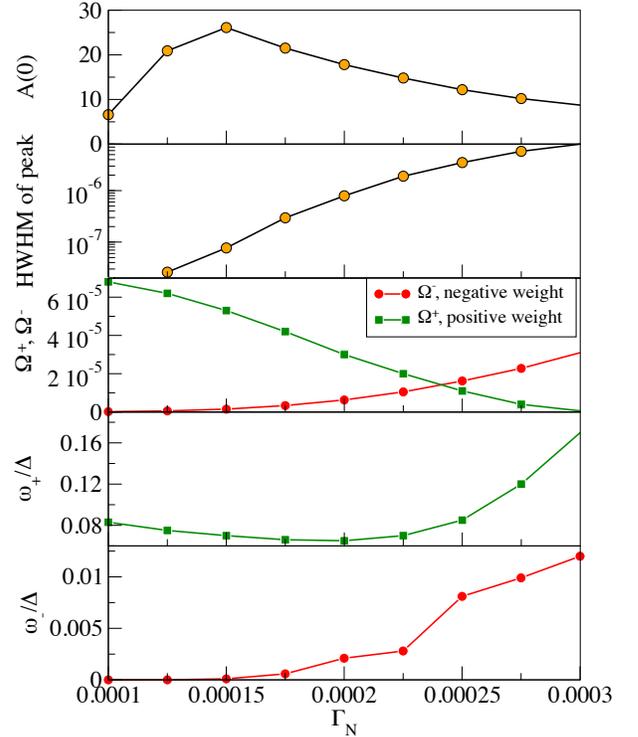}
	      \caption{(Color online). From top to bottom: zero-bias
	      peak in the spectral function
	        $A(0)$, width of the peak at zero frequency,
		$\Omega^{\pm}\equiv\int_{0}^{\Delta} d\omega
		B^{\pm}(\omega)$ (with $B^{\pm}(\omega)$ being the
		positive and
		negative parts of $B(\omega)$), and positive and
		negative peak
		positions of $B(\omega)$.}
		\label{figsup4}
		\end{figure}

One can also monitor the DS crossover via the anomalous spectral
function, as the peak position changes from positive to negative side,
indicating the occurence of the crossover. In Fig.~\textcolor{blue}{\ref{figsup3}(a)} we
have plotted the anomalous spectral function $B(\omega)$ when
$\Gamma_N$ is varied for a fixed value of $\Gamma_{SC}=0.004$, and
$U=0.05$ with $\Delta=2\times 10^{-4}$. For completeness we also
provide in Fig.~\textcolor{blue}{\ref{figsup3}(b)} the regular spectral function that
has a pronounced $\omega=0$ peak precisely when the anomalous spectral
function reverses sign. We note especially the case for
$\Gamma_N=2\times 10^{-4}$ [orange curve in Fig.~\textcolor{blue}{\ref{figsup3}(a)}].
The anomalous spectral function $B(\omega)$ has a complex behavior:
there is one positive peak at $\omega>0$, just like in the doublet
regime, but also one negative peak for $\omega>0$ (close to
$\omega=0$), just like in the singlet regime, so this is truly where the
crossover between the doublet and singlet 
regimes can be located. Note, however, that there are numerous
possible ways to define the ``crossover value'' of $\Gamma_N$:
zero-frequency spectral weight $A(0)$, crossing point of the
integrated weights of the anomalous spectral function $W_+$ and $W_-$,
or through peaks positions in $B(\omega)$.  The alternative crossover
values for $\Gamma_N$ attending to the previous criteria are
illustrated in Fig.~\textcolor{blue}{\ref{figsup4}}: the curves do not define
a unique special $\Gamma_N$ point.

We note that for $\Gamma_N=0$, the DS transition curve is determined
by the well-known $T_K=0.3 \Delta$ rule, where $T_K$ is the Kondo
temperature according to Wilson's definition, calculated for the SC
lead when the superconductivity is suppressed ($\Delta \to 0$) limit.
A relevant question is whether this rule still holds for $\Gamma_N
\neq 0$ with $T_K$ computed for $\Gamma_\mathrm{eff} =
\Gamma_N+\Gamma_{SC}$. We find that this produces the shift of the
cross-over line in the correct direction (toward smaller
$\Gamma_{SC}$; see Figs.~\textcolor{blue}{\ref{fig2}} and \textcolor{blue}{\ref{fig4}}), although quantitatively
we find that the effect of finite $\Gamma_N$ is more complex.

\section{The Schrieffer-Wolff transformation for a NS-impurity system}

\def \beq {\begin{equation}}
\def \edq {\end{equation}}
\def \bes {\begin{subequations}}
\def \eds {\end{subequations}}
\def \beqn {\begin{equation*}}
\def \edqn {\end{equation*}}
\def \nn  {\nonumber}
\def \dag {\dagger}
\def \Up {\Uparrow}
\def \Down {\Downarrow}
\def \up {\uparrow}
\def \down {\downarrow}
\def \eps {\epsilon}
\def \sm {\sigma}
\def \bsm {\bar{\sigma}}
\def \bk {\bar{k}}
\def \bs {\bar{s}}
\def \zhup {\sqrt{\frac{2\Gamma_{\up}}{\pi}}}
\def \zhdn {\sqrt{\frac{2\Gamma_{\down}}{\pi}}}
\def \veps {\varepsilon}
\def \calu {{\cal{U}}}
\def \calh {{\cal{H}}}
\def \call {{\cal{L}}}
\def \bcalh {\bar{\cal{H}}}
\def \bcall {\widetilde{\cal{L}}}
\def \calz {{\cal{Z}}}
\def \cald {{\cal{D}}}
\def \calg {{\cal{G}}}
\def \mG {{\mathscr{G}}}
\def \bcalg {\boldsymbol{\cal{G}}}
\def \calp {{\cal{P}}}
\def \calq {{\cal{Q}}}
\def \calf {{\cal{F}}}
\def \calo {{\cal{O}}}
\def \caln {{\cal{N}}}
\def \calt {{\cal{T}}}
\def \calr {{\cal{R}}}
\def \cals {{\cal{S}}}
\def \cali {{\cal{I}}}
\def \calm {{\cal{M}}}
\def \calw {{\cal{W}}}
\def \cale {{\cal{E}}}
\def \calc {{\cal{C}}}
\def \scrh {{\mathscr{H}}}
\def \scrg {{\mathscr{G}}}
\def \scrf {{\mathscr{F}}}
\def \scrs {{\mathscr{S}}}
\def \wcalh {\widetilde{\calh}}
\def \wPsi {\widetilde{\Psi}}
\def \hatb {\widehat{b}}
\def \hatc {\hat{c}}
\def \hatd {\hat{d}}
\def \hatf {\widehat{f}}
\def \hatz {\hat{z}}
\def \hatu {\hat{u}}
\def \hatL {\hat{L}}
\def \hatV {\hat{V}}
\def \hatw {\hat{\omega}}
\def \hatg {\widehat{g}}
\def \htau {\widehat{\tau}}
\def \hatgam {\widehat{\Gamma}}
\def \hatG {\widehat{{\cal{G}}}}
\def \hatQ {\widehat{{\cal{Q}}}}
\def \hatcP {\widehat{{\cal{P}}}}
\def \hatsig {\widehat{\Sigma}}
\def \hatsm {\widehat{\sigma}}
\def \bsig {{\mathbf{\Sigma}}}
\def \bV {{\mathbf{V}}}
\def \bfsm {{\boldsymbol{\sm}}}
\def \wtsig {{\widetilde{\Sigma}}}
\def \talpha {\widetilde{\alpha}}
\def \tell {\widetilde{\ell}}
\def \tmu {\widetilde{\mu}}
\def \tDelta {\widetilde{\Delta}}
\def \tp {\widetilde{p}}
\def \ty {\widetilde{y}}
\def \tz {\widetilde{z}}
\def \tPi {\widetilde{\Pi}}
\def \teps {\widetilde{\eps}}
\def \tom {\widetilde{\omega}}
\def \tgamma {\widetilde{\gamma}}
\def \tr {\text{Tr}}
\def \det {\text{det}}
\def \bo {\bf{1}}
\def \bt {\bar{t}}
\def \bp {\bar{p}}
\def \bq {\bar{q}}
\def \balpha {\bar{\alpha}}
\def \lAngle {\langle\langle}
\def \rAngle {\rangle\rangle}
\def \bzero {\tilde{0}}
\def \coffee {{\Huge{\Coffeecup}}}
\def \stop {{\Huge{\Stopsign}}}
\def \keyboard {{\Huge{\Keyboard}}}
\def \tGamma {\widetilde{\Gamma}}
\def \tPsi {\tilde{\Psi}}
\def \tg {\hat{\text{g}}}
\def \ttg {\text{g}}
\def \bttg {\text{$\mathbf{g}$}}
\def \mttg {\text{\bf{g}}}
\def \wtV {\widetilde{V}}
\def \wtt {\widetilde{t}}
\def \wtz {\widetilde{z}}
\def \tb {\widetilde{b}}
\def \tc {\tilde{c}}
\def \hatt {\hat{t}}
\def \bA {\mathbf{A}}
\def \bX {\mathbf{X}}
\def \bY {\mathbf{Y}}
\def \bnu {\boldsymbol{\nu}}
\def \matone {\mathbb{1}}
\def \Pf {\text{Pf}}
\def \whand {\huge{\Writinghand}}
\def \hc {\text{H.c.}}

\providecommand{\abs}[1]{\lvert#1\rvert}
\providecommand{\norm}[1]{\lVert#1\rVert}
\providecommand{\brakets}[2]{\langle#1|#2|#1\rangle}
\providecommand{\Brakets}[3]{\langle#1|#2|#3\rangle}
\providecommand{\nbraket}[1]{\left\langle#1\right\rangle}
\providecommand{\dbraket}[1]{\langle\langle#1\rangle\rangle}

We perform here the Schrieffer-Wolff transformation
\textcolor{blue}{\cite{Schrieffer66}} for the NS-impurity system.  By doing this we
obtain the exchange couplings for the impurity spin-flip processes
from which a functional form for the Kondo temperature can be
inferred. Our starting point is a hybrid normal-superconductor
Anderson Hamiltonian \beq \calh = \calh_N + \calh_S + \calh_D +
\calh_T = \calh_0 + \calh_T\,, \edq where \bes
\begin{align}
\calh_N &= \sum_{k,\sm} \eps_{k_N} c_{k_N\sm}^{\dag}c_{k_N\sm}\,, \\
\calh_S &= \sum_{k,\sm} \eps_{k_{SC}}
c_{k_{SC}\sm}^{\dag}c_{k_{SC}\sm} + \sum_k \left(\Delta
c_{k_{SC}\up}^{\dag}c_{\bk_{SC}\down}^{\dag} + \hc\right)\,,\\
\calh_D &= \sum_{\sm} \eps_{d\sm} d_{\sm}^{\dag}d_{\sm} +
Un_{d\up}n_{d\down}\,, \\
\calh_T &= \sum_{\alpha,k,\sm} \left( V_{\alpha}
c_{k_{\alpha}\sm}^{\dag} d_{\sm} + \hc\right)\,.
\end{align}
\eds
The operator $c_{k_{\alpha}\sm}$ ($c_{k_{\alpha}\sm}^{\dag}$)
annihilates (creates) an electron with
wave-vector $k$ ($\bk = -k$), energy $\epsilon_{k_\alpha}$ and spin
$\sm=\{\up,\down\}$ in the
normal or superconducting lead ($\alpha=\{ N,SC \}$). Similarly,
$d_{\sm}$ ($d_{\sm}^{\dag}$) destroys (creates) an electron with spin
$\sigma$ and energy
$\epsilon_{d\sm}$ at the impurity level. $n_{d\sm}=d_{\sm}^\dag
d_{\sm}$ 
is the impurity occupation and
$U$ denotes the on-site Coulomb interaction. Tunneling amplitudes for
normal-impurity 
and superconducting-impurity processes are indicated by $V_{N}$, and
$V_{S}$, respectively.  
Here, $\Delta$ denotes the superconducting gap considered to be real.

It is convenient to introduce the Bogoliubov-Valatin transformation 
\textcolor{blue}{\cite{Bogoliubov58A, Bogoliubov58B, Valatin58}}
\beq
\begin{pmatrix}
c_{k_{SC}\up}
\\
c_{\bk_{SC}\down}^{\dag}
\end{pmatrix}
=
\begin{pmatrix}
u_k & -v_k \\
v_k & u_k
\end{pmatrix}
\begin{pmatrix}
a_{k}
\\
b_{\bk}^{\dag}
\end{pmatrix}\,.
\edq
The superconducting coherence factors satisfy the relations
\beq
u_k^2 = \frac{1}{2}\left(1+\frac{\eps_{k_{SC}}}{E_k}\right), \,\,
v_k^2 = \frac{1}{2}\left(1-\frac{\eps_{k_{SC}}}{E_k}\right)
\label{eq:uv}
\edq
with $E_k = \sqrt{\eps_{k_{SC}}^2 + \Delta^2}$.
$u_k = u_{\bk}$, $v_k = v_{\bk}$, and $E_k = E_{\bk}$ are obeyed.  
Using the transformation, $\calh_S$ becomes
\beq
\calh_S = \sum_k E_k(a_k^{\dag}a_k + b_k^{\dag}b_k) \,,
\edq
while $\calh_T$ is expressed in the form
\begin{multline}
\calh_T = \sum_k \left\{ V_S\left[\left(u_k a_k^{\dag} - v_k
b_{\bk}\right)d_{\up} + \left(v_k a_k + u_k
b_{\bk}^{\dag}\right)d_{\down}\right]
\right. \\ \left.
+ V_S \left[d_{\up}^{\dag}\left(u_k a_k - v_k b_{\bk}^{\dag}\right) +
d_{\down}^{\dag}\left(v_k a_k^{\dag} + u_k b_{\bk}\right)\right]\right\}
\\
+ \sum_{k,\sm} V_N\left(c_{k_N\sm}^{\dag}d_{\sm} +
d_{\sm}^{\dag}c_{k_N\sm}\right) \,.
\end{multline}

We make an unitary transformation to get an effective Hamiltonian
\begin{multline}
\calh_{eff} = e^{S} \calh e^{-S} = \sum_{n=0}^{\infty}
\frac{1}{n!}[S,\calh]_n \approx 
\\ 
\calh_0 + \calh_T + [S,\calh_0] + [S,\calh_T]
+\frac{1}{2!}[S,[S,\calh_0]]\,,
\end{multline}
where $[S,\calh]_0 = \calh$ and $[S,\calh]_n =
[S,[S,[\cdots,[S,\calh]]\cdots]]$.
Our purpose is to find an $S$ which satisfies
\beq
\calh_T + [S,\calh_0] = 0
\label{eq:Scond}
\edq
The effective Hamiltonian then becomes
\beq
\calh_{eff} = \calh_0 + [S,\calh_T] + \frac{1}{2!}[S,-\calh_T] =
\calh_0 + \frac{1}{2}[S,\calh_T]\,.
\edq
For our setup, the generator $S=S_0-S_0^\dagger$ reads \textcolor{blue}{\cite{Salomaa88}}
\begin{widetext}
\begin{multline}
S_0 = \sum_k V_S \left\{\left[\frac{n_{d\down}}{E_{k} - \eps_{d\up} -
U} + \frac{1-n_{d\down}}{E_{k} - \eps_{d\up}}\right]u_ka_k^{\dag}d_{\up}
+ \left[\frac{n_{d\down}}{E_{\bk} + \eps_{d\up} + U} +
\frac{1-n_{d\down}}{E_{\bk} + \eps_{d\up}}\right]v_kb_{\bk}d_{\up}
\right. \\ \left.
- \left[\frac{n_{d\up}}{E_{\bk} + \eps_{d\down} + U} +
\frac{1-n_{d\up}}{E_{\bk} + \eps_{d\down}}\right]v_ka_{\bk}d_{\down}
+ \left[\frac{n_{d\up}}{E_{k} - \eps_{d\down} - U} +
\frac{1-n_{d\up}}{E_{k} - \eps_{d\down}}\right]u_kb_{k}^{\dag}d_{\down}
\right\}
\\
+ \sum_{k,\sm} V_N\left[\frac{n_{d\bsm}}{\eps_{k_N} - \eps_{d\sm} - U}
+ \frac{1-n_{d\bsm}}{\eps_{k_N} -
\eps_{d\sm}}\right]c_{k_N\sm}^{\dag}d_{\sm}
\end{multline}
\end{widetext}
where $\bsm = \down/\up$ for $\sm = \up/\down$. 
It is easy to check that the generator $S$ satisfies
Eq.~\textcolor{blue}{\eqref{eq:Scond}}.

The transformed Hamiltonian can be arranged in a concise form
\beq
\calh_{eff} = \calh_0' + \calh_{PS} + \calh_{SF} + \calh_{QSF} \,.
\edq
Here, $\calh_0'$ corresponds to $\calh_0$ with renormalized parameters
and $\calh_{PS}$ denotes the potential scattering of electrons off the
impurity.
The impurity-electron spin-flip processes are described by
\begin{multline}
\calh_{SF} = -\frac{1}{2}\sum_{k,p}\sum_{\sm}
J_{N,N,k,p}c_{k_N\sm}^{\dag}c_{p_N\bsm}d_{\bsm}^{\dag}d_{\sm}
\\
-\frac{1}{2}\sum_{k,p}\sum_{\sm}
J_{N,S,k,p}c_{k_N\sm}^{\dag}c_{p_{SC}\bsm}d_{\bsm}^{\dag}d_{\sm}
\\
-\frac{1}{2}\sum_{k,p}\sum_{\sm}
J_{S,N,k,p}c_{k_{SC}\sm}^{\dag}c_{p_N\bsm}d_{\bsm}^{\dag}d_{\sm}
\\
-\frac{1}{2}\sum_{k,p}\sum_{\sm}
J_{S,S,k,p}c_{k_{SC}\sm}^{\dag}c_{p_{SC}\bsm}d_{\bsm}^{\dag}d_{\sm}
\\
+ \frac{1}{2}\sum_{\alpha}\sum_{k,p}\sum_{\sm} {\rm sgn}(\sm)
\left(T_{S,\alpha,k,p}
c_{p_{\alpha}\bsm}c_{\bk_{SC}\bsm}d_{\bsm}^{\dag}d_{\sm} + \hc\right)\,,
\end{multline}
where
\begin{multline}
J_{N,N,k,p} = V_N^2\left[\frac{1}{\eps_{k_N}-\eps_d-U} -
\frac{1}{\eps_{k_N}-\eps_d}
\right. \\ \left.
+ \frac{1}{\eps_{p_N}-\eps_d-U} - \frac{1}{\eps_{p_N}-\eps_d}\right] \,,
\end{multline}
\begin{multline}
J_{N,S,k,p} = J_{S,N,p,k} = V_NV_S\left[\frac{1}{\eps_{k_N}-\eps_d-U}
- \frac{1}{\eps_{k_N}-\eps_d}\right] 
\\
+ V_SV_N\Big[\frac{u_p^2}{E_{p}-\eps_d-U} - \frac{u_p^2}{E_{p}-\eps_d}
\\ 
-\frac{v_p^2}{E_{\bp}+\eps_d+U} + \frac{v_p^2}{E_{\bp}+\eps_d}\Big]\,, 
\end{multline}


\begin{multline}
J_{S,S,k,p} = V_S^2\Big[\frac{u_k^2}{E_{k}-\eps_d-U} -
\frac{u_k^2}{E_{k}-\eps_d} \\ 
-\frac{v_k^2}{E_{\bk}+\eps_d+U} + \frac{v_k^2}{E_{\bk}+\eps_d}
\\
+ \frac{u_p^2}{E_{p}-\eps_d-U} - \frac{u_p^2}{E_{p}-\eps_d} \\ 
-\frac{v_p^2}{E_{\bp}+\eps_d+U} + \frac{v_p^2}{E_{\bp}+\eps_d}\Big]\,, 
\label{eq:JSS}
\end{multline}


\begin{multline}
T_{S,\alpha,k,p} = V_SV_{\alpha} u_kv_k\Big[\frac{1}{E_k-\eps_d-U} -
\frac{1}{E_k-\eps_d} 
\\+ \frac{1}{E_{\bk}+\eps_d+U}-\frac{1}{E_{\bk}+\eps_d}\Big]\,. 
\end{multline}
The final term shows the charge-transfer interaction given by
\begin{multline}
\calh_{QSF} =
-\frac{1}{2}\sum_{\alpha}\sum_{k,p}\sum_{\sm}\left(K_{N,\alpha,k,p}c_{k_N\sm}^{\dag}c_{p_{\alpha}\bsm}^{\dag}d_{\bsm}d_{\sm}
+ \hc\right)
\\
-\frac{1}{2}\sum_{\alpha}\sum_{k,p}\sum_{\sm}\left(K_{S,\alpha,k,p}c_{k_{SC}\sm}^{\dag}c_{p_{\alpha}\bsm}^{\dag}d_{\bsm}d_{\sm}
+ \hc\right)
\\
+ \frac{1}{2}\sum_{\alpha}\sum_{k,p}\sum_{\sm} {\rm sgn}(\sm)
\left(L_{S,\alpha,k,p}
c_{p_{\alpha}\bsm}^{\dag}c_{\bk_{SC}\bsm}d_{\bsm}d_{\sm} + \hc\right)\,,
\end{multline}
where
\bes
\beq
K_{N,\alpha,k,p} = V_NV_{\alpha} \left[\frac{1}{\eps_{k_N}-\eps_d-U} -
\frac{1}{\eps_{k_N}-\eps_d}\right]\,,
\edq
\begin{multline}
K_{S,\alpha,k,p} = V_SV_{\alpha}\Big[\frac{u_k^2}{E_{k}-\eps_d-U} -
\frac{u_k^2}{E_{k}-\eps_d} 
\\ - \frac{v_k^2}{E_{\bk}+\eps_d+U} +
\frac{v_k^2}{E_{\bk}+\eps_d}\Big]\,,\nonumber
\end{multline}
\begin{multline}
L_{S,\alpha,k,p} = V_SV_{\alpha} u_kv_k\Big[\frac{1}{E_k-\eps_d-U} -
\frac{1}{E_k-\eps_d} 
\\+ \frac{1}{E_{\bk}+\eps_d+U}-\frac{1}{E_{\bk}+\eps_d}\Big]\,.
\nonumber
\end{multline}
\eds
Since double occupation of the impurity site is suppressed for $U>0$, 
usually $\calh_{QSF}$ is neglected \textcolor{blue}{\cite{Schrieffer66, Salomaa88}}.

We focus on the spin-flip exchange interactions responsible for the
occurrence of Kondo effect. 
First, for the normal spin-flip exchange constant 
$J_{N,N,k,p}$ it can be approximated as
\beq
J_{N,N,k,p} \approx 2V_N^2 \frac{U}{(\eps_d+U)\eps_d} \,.
\edq
Second, by inserting Eqs.~\textcolor{blue}{\eqref{eq:uv}} into \textcolor{blue}{\eqref{eq:JSS}} the
exchange constant $J_{S,S,k,p}$ mediated by the superconducting lead
reads
\begin{multline}
J_{S,S,k,p} = \frac{V_S^2}{2}\Big[\frac{U}{(E_k-\eps_d-U)(E_k-\eps_d)} 
\\+ \frac{U}{(E_k+\eps_d+U)(E_k+\eps_d)} \Big]
\\
+
\frac{V_S^2}{2}\cdot\frac{\eps_{k_{SC}}}{E_k}\Big[\frac{U}{(E_k-\eps_d-U)(E_k-\eps_d)}
\\- \frac{U}{(E_k+\eps_d+U)(E_k+\eps_d)} \Big] + (k\leftrightarrow p) \,.
\end{multline}
Notice that for $\Delta\to 0$ we recover the exchange constant
equivalent to the normal lead 
\beq
J_{S,S,k,p} \approx 2V_S^2 \frac{U}{(\eps_d+U)\eps_d} \,. 
\edq
In addition, it is worth to realize that at the particle-hole
symmetric point ($U=-2\veps_d$) $J_{S,S,k,p}$ can be simplified to
\beq
J_{S,S,k,p} = V_S^2\left[\frac{U}{E_k^2-U^2/4}\right] +
(k\leftrightarrow p) \,.
\edq
Thus, if $\Delta \ll U$ we also recover the normal lead limit, i.e.
$J_{S,S,k,p} \approx -8V_S^2/U$.
On the other hand, in the limit of $\Delta \gg U$, $J_{S,S,k,p}$ can
be neglected.
The exchange couplings mediated by both the superconducting and normal
leads are described by $J_{N,S,k,p}$ and $J_{S,N,k,p}$.
Similar to $J_{S,S,k,p}$, at the particle-hole symmetric point it
reduces to
\beq
J_{N,S,k,p} = J_{S,N,p,k} \approx -\frac{4V_NV_S}{U} +
V_SV_N\left[\frac{U}{E_p^2-U^2/4}\right] \,.
\edq
We notice that the second term can be neglected in the limit of
$\Delta \gg U$.
Together with vanishing of $J_{S,S,k,p}$, this partially explains why
we observe the needle Kondo peak in the doublet regime.
Finally, the constant $T_{S,\alpha,k,p}$ manifests itself only when
the superconducting lead is present since it is proportional to
$u_kv_k \propto \Delta$.
Also, observe that $T_{S,\alpha,k,p}$ vanishes at the particle-hole
symmetric point.

\begin{figure}
  \centering \includegraphics[width=0.5\textwidth]{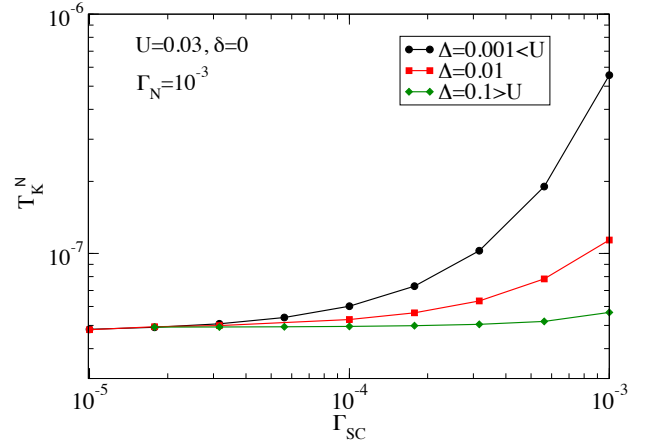}
  \caption{(Color online) NRG results for the Kondo temperature
  $T_K^{N}$ of the needle resonance as a function of the exchange
  coupling to the superconducting lead, $\Gamma_{SC}$, for several
  values of the BCS gap $\Delta$, both in the small $\Delta$ and large
  $\Delta$ limits.}
  \label{figsup5}
  \end{figure}

We may contrast these results with the work based on the continuous
unitary transformation (CUT) \textcolor{blue}{\cite{Zapalska:13}}, which is essentially a
continuous version of the Schrieffer-Wolff transformation. That work
was done in the $\Delta \to \infty$ limit, resulting in the effective
Kondo exchange coupling constant $J = -4U|V_N|^2/(U^2+4\Delta_d^2)$,
where $\Delta_d$ is the proximity-induced on-dot pairing $\Delta_d =
\Gamma_{SC}/2$. This implies that with increasing coupling to the SC
lead the exchange coupling grows weaker. That results is not general,
however: it holds only in the limit of $\Delta\to\infty$. At the Fermi
level, we find more generally (for $\epsilon_d=-U/2$):
\beq
\begin{split}
J_{NN} &= -\frac{8V_N^2}{U}, \\
J_{SS} &= \frac{2V_S^2 U}{\Delta^2-U^2/4}, \\
J_{NS} = J_{SN} &= -V_N V_S \left( \frac{4}{U} +
\frac{U}{U^2/4-\Delta^2} \right).
\end{split}
\edq
For $V_S \ll V_N$, the leading effect is that of the mixed term
$J_{NS}$, since $J_{SS}$ is subleading in $V_S$. For small $\Delta$,
the expression between the parenthesis is positive, thus finite $V_S$
leads to an {\it enhancement} of the exchange coupling. This is also
explicity confirmed by our numerical NRG results in the $\Delta < U$
limit even for $V_S \sim V_N$, see Fig.~\textcolor{blue}{\ref{figsup5}}. In fact, the
numerical results indicate an enhancement of $T_K$ even for large
$\Delta$ approaching the half-bandwidth $D=1$. 

\section{$\Gamma_N$ dependence of $T_K^N$}  
  
\begin{figure}
    \centering \includegraphics[width=0.5\textwidth]{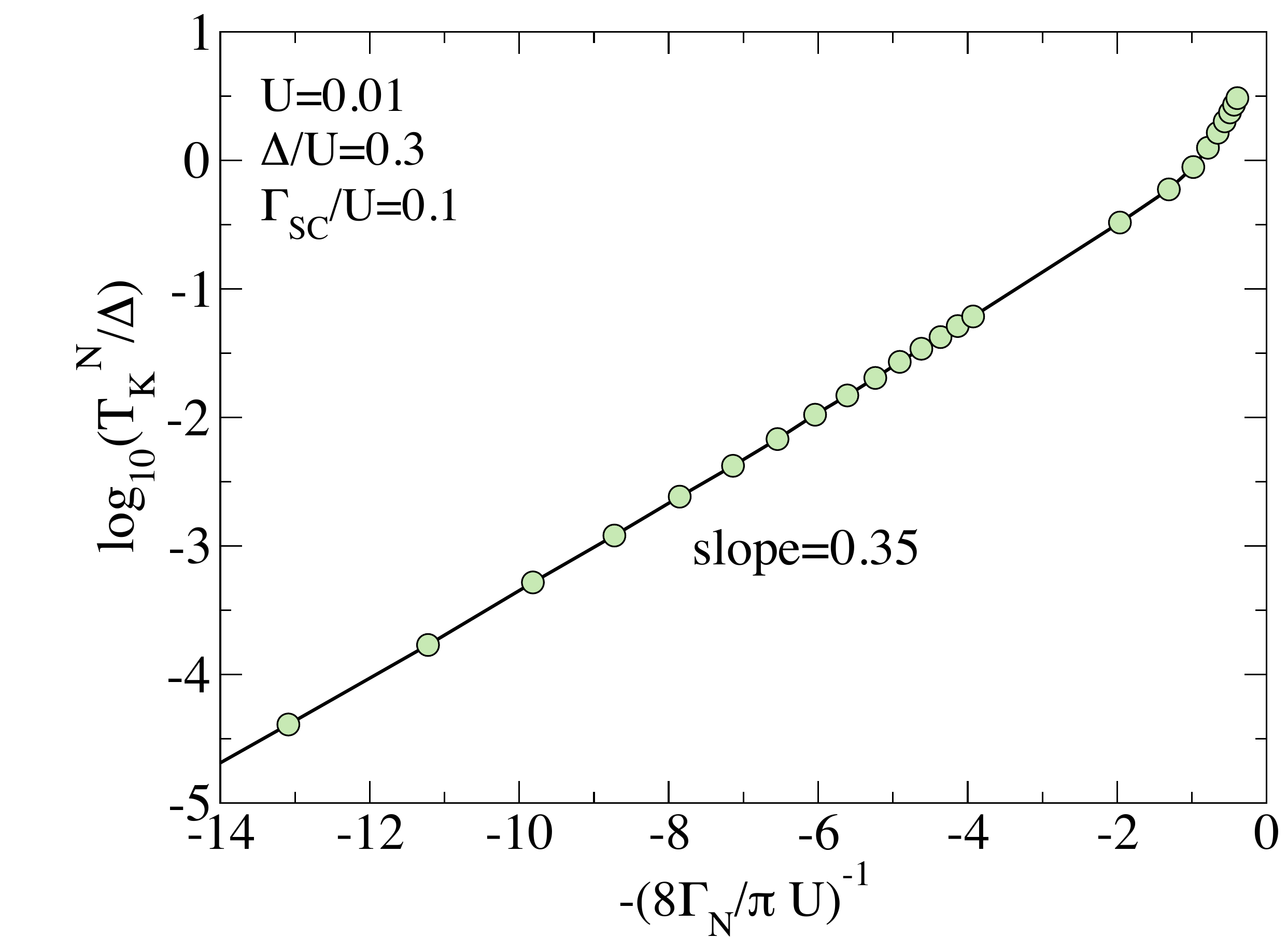}
    \caption{(Color online) NRG results for the Kondo temperature
    $T_K^{N}$ of the needle resonance as a function of the exchange
    coupling to the normal lead, $\Gamma_{N}$.}
    \label{figsup6}
    \end{figure}

  The $\Gamma_N$ dependence of the Kondo temperature $T_K^N$ is shown
  in Fig.~\textcolor{blue}{\ref{figsup6}}. The behaviour for small $\Gamma_N$ is
  exponential, but with a non-standard factor in the exponent:
  \begin{equation}
  T_K^N \propto \exp\left(-c \frac{\pi U}{8\Gamma_N} \right),
  \end{equation}
  where $c$ is a constant of order 1 which depends on $\Delta/U$ and
  $\Gamma_{SC}/U$ ratios; for parameters in the plot, we find $c=0.35$.
  For the standard single-impurity Anderson model with normal lead only,
  $c=1$. The deviation from $c=1$ (towards smaller values) indicates a
  renormalization of the charge fluctuation scale $U$ by the coupling to
  the superconducting lead. $c$ decreases ($U$ renormalizes more
  significantly) with increasing $\Gamma_{SC}$ and decreasing $\Delta$.
  
\end{document}